\documentclass[graybox]{svmult}

\usepackage{amsfonts}
\usepackage{amsmath}
\usepackage{amssymb}
\usepackage{mathptmx}       
\usepackage{helvet}         
\usepackage{courier}        
\usepackage{type1cm}        
\usepackage{makeidx}         
\usepackage{graphicx}        
\usepackage{multicol}        
\usepackage[bottom]{footmisc}

\makeindex             

\usepackage{dsfont}
\usepackage{url}

\newcommand{\EE}{\mathbb{E}}

\newcommand{\MM}{\mathbb{M}}

\newcommand{\PP}{\mathbb{P}}
\newcommand{\QQ}{\mathbb{Q}}
\newcommand{\RR}{\mathbb{R}}

\newcommand{\Za}{ {\cal Z }}

\newcommand{\point}{\mbox{\LARGE .}}

\def \PP{\mathbb{P}}
\def \RR{\mathbb{R}}
\def \EE{\mathbb{E}}
\def \QQ{\mathbb{Q}}

\begin{document}

\title*{An introduction to stochastic particle integration methods: with applications to risk and insurance}
\author{P. Del Moral, G. W. Peters, Ch. Verg\'e}
\institute{P. Del Moral \at INRIA Research Director, INRIA, Bordeaux Mathematical Institute, Bordeaux-Sud Ouest,\\
Professeur chargé de cours Ecole Polytechnique (CMAP)\\
 \email{Pierre.Del_Moral@inria.fr}
\and G.W. Peters \at Department of Statistical Science, University College London;\\
School of Mathematics and Statistics, UNSW; \\
Commonwealth Scientific and Industrial Research Organization, CSIRO-CMIS,\\
 \email{gareth.peters@ucl.ac.uk}}
%
%
\maketitle

\vspace{-2cm}
\abstract{This article presents a guided introduction to a general class of interacting particle methods and explains throughout how such methods may be adapted to solve general classes of inference problems encountered in actuarial science and risk management. Along the way, the resulting specialized Monte Carlo solutions are discussed in the context of how they compliment alternative approaches adopted in risk management, including closed from bounds and asymptotic results for functionals of tails of risk processes.\\
The development of the article starts from the premise that whilst interacting particle methods are increasingly used to sample from complex and high-dimensional distributions, they have yet to be generally adopted in inferential problems in risk and insurance. Therefore, we introduce in a principled fashion the general framework of interacting particle methods, which goes well beyond the standard particle filtering framework and Sequential Monte Carlo frameworks to instead focus on particular classes of interacting particle genetic type algorithms. These stochastic particle integration techniques can be interpreted as a universal acceptance-rejection sequential particle sampler equipped with adaptive and interacting recycling mechanisms which we reinterpret under a Feynman-Kac particle integration framework. These functional models are natural mathematical extensions of the traditional change of probability measures, common in designing importance samplers.\\
Practically, the particles evolve randomly around the space independently and to each particle is associated a positive potential function. Periodically,  particles with high potentials duplicate at the expense of low potential particle which die. This natural genetic type selection scheme appears in numerous applications in applied probability, physics, Bayesian statistics, signal processing, biology, and information engineering. It is the intention of this paper to introduce them to risk modeling. 
}

\section{Introduction to Stochastic Particle Integration}
\label{sec:1}
The intention of this paper is to introduce a class of stochastic particle based integration techniques to a broad community, with a focus on risk and insurance practitioners, who we believe will benefit from the development of such inferential methods in several problems encountered in these domains. A key motivation for this endeavor is due to the fact that over the last two decades, stochastic particle integration models have been extensively and routinely used in engineering, statistics and physics under sometimes different names, such as: particle filters, bootstrap or genetic filters, population Monte Carlo methods, sequential Monte Carlo models, genetic search models, branching and multi-level splitting particle rare event simulations, condensation models, go-with-the winner, spawning models, walkers population reconfigurations, pruning-enrichment strategies, quantum and diffusion Monte Carlo, rejuvenation models, and many others. They have however not yet been routinely applied to develop solutions in important financial domains such as those we discuss in this tutorial type review, in which we discuss and illuminate areas they will benefit the fields of risk and actuarial science. To contain the scope of such an endeavor we focus the application domain discussions on the widely used class of problems based around the consideration of single risk processes under the Loss Distributional Approach, see detailed discussions in \cite{peters2006bayesian}, \cite{peters2009sequential},\cite{shevchenko2009implementing} and in the dynamic setting \cite{peters2009dynamic} and \cite{lambrigger2007quantification}.

We begin with the introduction of the fundamental background for interacting particle systems highlighting key papers in their developments through a range of different science disciplines, before introducing aspects of these stochastic methods to risk and insurance. It is important that practitioners when first encountering such Feynman-Kac interacting particle methods are aware that they encompass a far more general gamut of stochastic integration and optimization methodologies then the most well known sub-class of such methods knows as particle filters, which are typically utilized in inference under latent process state space models in engineering and statistics. It is the intention of this article to explain the key papers and ideas in a far more general framework which is much more encompassing than the special subset of particle filter based algorithms. It is for this reason that me mention that the stochastic methods discussed in this paper are applicable to a significantly larger subset of problems than the standard particle filter approach that may be thought of when first encountering Sequential Monte Carlo.

Then we proceed through a selection of key features of their development, focusing on a sub-class of such methods of relevance to the application domain explored in this manuscript, risk and insurance. The origins of stochastic particle simulation certainly starts with the seminal paper of N. Metropolis and S. Ulam~\cite{metro}. As explained by these two physicists in the introduction of their pioneering article, the Monte Carlo method is, "essentially, a statistical approach to the study of differential equations, or more generally, of integro-differential equations that occur in various branches of the natural sciences". The links between genetic type particle Monte Carlo models and quadratic type parabolic integro-differential equations has been developed in the beginning of 2000' in the series of articles on continuous time models~\cite{dm-miclo1,dm-miclo2}.

The earlier works on heuristic type genetic particle schemes seem to have started in Los Alamos National Labs with works of M.N. Rosenbluth and A.W. Rosenbluth~\cite{rosen}, and T.E. Harris and H. Kahn~\cite{harris}. We also quote the work on artificial life of Nils Aall Barricelli~\cite{barri1,barri2}. In all of these works, the genetic Monte Carlo scheme is always presented as a natural heuristic resampling type algorithm to generate random population models, to sample molecular conformations, or to estimate high energy particle distributions, without a single convergence estimate to ensure the performance, nor the robustness of the Monte Carlo sampler. 

The mathematical foundations, and the performance analysis of all of these discrete generation particle models are rather recent. The first rigorous study in this field seems to be the article~\cite{dmp96} published in 1996 on the applications of particle methods to nonlinear estimation problems. This article provides the first proof of the unbiased property of particle likelihood approximation models (lemma 3 page 12); and adaptive resampling criteria w.r.t. the weight dispersions (see remark 1 on page p.4).  We also quote the first articles presenting heuristic type particle filters ~\cite{gsm93,k93}, and a series of earlier research reports~\cite{dmp-c1,dmp-c2,dmp-c3,dmp-c4}.

For an in depth description of the origins of particle methods, and their applications we refer to the following studies~\cite{dmp98,dpr11}. These articles also contain new stochastic models and methods including look-ahead type strategies (section 4.2.2), reducing the variance using conditional explorations w.r.t. the observation sequences (example 3 p. 40), local errors transport models (see the proof of theorem 1 on page 11), mean field models w.r.t. the occupation measures of random trees (section 3.2). 

A more detailed review of particle models in discrete and continuous time can be found in~\cite{dm00,dm04}. In the research monograph the reader will find a detailed discussion on particle models and methods including acceptance-rejection with recycling particle strategies, interacting Kalman filters a.k.a. Rao-Blackwellized particle filters (section 2.6, and section 12.6.7), look-ahead type strategies (section 12.6.6), genealogical tree models and branching strategies (section 11), and interacting Metropolis-Hasting models (chapter 5).

The practitioner will find in the research books~\cite{dm04,dm12} a source of useful convergence estimates as well as a detailed list of concrete examples of particle approximations for real models, including restricted Markov chain simulations, random motions in absorbing media, spectral analysis of Schrodinger operators and Feynman-Kac semigroups, rare event analysis, sensitivity measure approximations, financial pricing numerical methods, parameter estimation in HMM models, island particle models, interacting MCMC models, statistical machine learning, Bayesian inference, Dirichlet boundary problems, nonlinear filtering problems, interacting Kalman-Bucy filters, directed polymer simulations, stochastic optimization, and interacting Metropolis type algorithms. 

There is an extensive number of texts on particle simulation and sequential Monte Carlo samplers, many of them contain much practically oriented discussions including Bayesian inference, nonlinear filtering and optimization, as well as optimal control problems. For a further discussion on the origins and the applications of these stochastic models, we refer the reader to the following texts ~\cite{aru,cappe,cappe2,ddj-06,petersMSc,dmkm,Doucet:2001,DoucetJohansen:2008,kantas,k96,najim,peters2009sequential}, and the references therein. 

Particle methods are yet to be routinely or widely introduced to areas of risk and insurance modeling. The initial examples that have been developed are detailed in \cite{peters2007simulation}, where a special sub-class of such methods was developed for an important set of risk management problems. It is the intention of this paper to highlight aspects of this class of problems and the stochastic particle solutions. 

Before, proceeding with the introduction of stochastic particle methods, we first provide an overview of the domain of application to be considered in risk and insurance and a mathematical discussion that will motivate the relevance of such stochastic particle integration methods. 

\section{Motivation for Stochastic Particle Solutions:\\ \textit{How Such Methods May Complement Risk Process Asymptotics}}
\label{sec:1}
Here we provide motivation to explain how and why risk management and actuarial sciences can benefit from the development of interacting particle system inferential solutions to an important subset of generic problems faced by practitioners in these domains. In particular we focus on a an aspect of single risk loss processes described under a Loss Distributional Approach (hereafter LDA) framework, see discussion in \cite{nevslehova2006infinite}, \cite{shevchenko2009implementing}, \cite{peters2009dynamic} and the books \cite{klugman1998loss} and \cite{shevchenko2011modeling} for the background on such modeling approaches in risk. For basic discussions on how such problems relate to a large class of non-life insurance problems see examples in \cite{peters2009sequential}. 

\subsection{The Loss Distributional Approach and Risk Management: \;\;\;\; \;\;\;\;  a tale of light to heavy tails}
In this section we fist motivate and introduce the context of LDA modeling in risk and insurance. Then we present three key challenges associated with working with such models faced by risk and insurance practitioners, thereby effectively detailing important inference challenges faced by such practitioners. Next, we provide a brief specifically selected survey of closed form analytic results known in the actuarial and risk literature for sub-classes of such LDA models as the Single Loss Approximations (hereafter SLA). We detail the closed form solution for the light-tailed severity distribution case and then explain how such approaches can not be obtained in such a form in the heavy-tailed sub-exponential risk process settings, often of interest in the domain of risk and insurance. As a result, we briefly present the results recently developed in actuarial literature for the heavy tailed case corresponding to the first order and second order asymptotic approximations, see comprehensive discussions in a general context in \cite{albrecher2010higher}, \cite{degen2010calculation} and the books, \cite{barbe2009asymptotic} and the forthcoming \cite{CruzPetersShevchenko}. 

We conclude this section by observing that according to regulatory standards and indeed good risk management practice \textit{such approximations are often required to be accompanied with numerical and statistical solutions which can more readily take into account model uncertainty, parameter uncertainty and the fact that such approximations are inherently asymptotic in nature, and may be inaccurate outside of the neighborhood of infinity.} In this regard we summarize a class of interacting particle solutions recently developed to address such estimations which complement such closed form asymptotic results. 

Consider the widely utilized insurance model known as a single risk Loss Distributional Approach model. This represents the standard under the Basel II/III capital accords \cite{basel2001basel} and involves an annual loss in a risk cell (business line/event type) modeled as a compound distributed random variable,
\begin{equation}
Z_{t}^{\left( j\right) }=\sum\limits_{s=1}^{N_{t}^{\left( j\right)
}}X_{s}^{\left( j\right) }\left( t\right),  
\label{AnnLoss1}
\end{equation}
for $t=1,2,\ldots,T$ discrete time (in annual units) and index $j$ identifies the risk cell. Furthermore, the annual number of losses is denoted by $N_{t}^{(j)}$ which is a random variable distributed according to a frequency counting distribution $P^{(j)}(\cdot) $, typically Poisson, Binomial or Negative Binomial and the severities (losses) in year $t$ are represented by random variables $X_{s}^{(j)}(t)$, $s \ge 1$, distributed according to a severity distribution $F^{(j)}(\cdot)$.

In constructing this model we assume that all losses are i.i.d. with $X_{s}^{\left( j\right) }\left( t\right) \sim F_X(x)$ and that the severity distribution is continuous with no atoms in the support $[0,\infty)$. As a consequence, linear combinations (aggregation) of losses in a given year, denoted by 
$$
S(t,n) = \sum_{s=1}^n X_{s}^{\left( j\right) }\left( t\right) \sim F_S(x)
$$ 
have the following analytic representation:
\begin{equation*}
\begin{split}
F_S(x) &= \left(F \star F \star \cdots F\right)(x) = \int_{[0,\infty)}F^{(n-1)\star}(x-y)dF(x).
\end{split}
\end{equation*}
We also observe that due to a result in \cite{feller1966introduction} if $F(x)$ has no atoms in $[0,\infty)$ then the n-fold convolution of such severity distributions will also admit no atoms in the support $[0,\infty)$. The implications of this for such Interacting Particle based numerical procedures (IS, SMC, MCMC) is that it ensures numerical techniques are well defined for such models when considering ratios of densities on the support $[0,\infty)$. In addition we note that continuity and boundedness of severity distribution $F_X(x)$ is preserved under $n$-fold convolution $\Rightarrow$ if $F_X(x)$ admits a density $\frac{d}{dx}F_X(x)$ then so does $F_S(x)$. For most models such analytic representations of the combined loss distribution are non closed form, with the exception of special sub-families of infinitely divisible severity distribution models, see \cite{peters2011analytic}.

It is important in practice to consider carefully the single risk processes in which business managers believe will produce infrequent losses with very high consequence. Modeling such risk processes typically requires sub-exponential severity distributions. If one considers losses $X_1,\ldots,X_n,\ldots$ as independent positive random variables with distribution $F(x) = \mathbb{P}\left(X_k < x\right), \; \; \forall k \in \left\{1,2,\ldots,n,\ldots \right\}$. Then the class of sub-exponential distributions $(F(x) \in \mathcal{F})$ satisfy the limits
\begin{equation}
\lim_{x \rightarrow \infty} \frac{1-F^{n\star}(x)}{1-F(x)}=n
\end{equation}
if and only if
\begin{equation}
\lim_{x \rightarrow \infty} \frac{1-F^{2\star}(x)}{1-F(x)}=2.
\end{equation}
The sub-exponential family of distributions $\mathcal{F}$ defines a class of heavy-tailed severity models in which \cite{pitman1980subexponential} demonstrated the necessary and sufficient condition for membership being, for a severity distribution $F \in \mathcal{F}$ if and only if the tail distribution $\overline{F}(x) = 1-F(x)$ satisfies 
$$
\lim_{x \rightarrow \infty} \int^{x}_0 \frac{\overline{F}(x-y)}{\overline{F}(x)}F(y)dy=1.
$$
Alternatively one may characterize the family of distributions $F \in \mathcal{F}$ by those that satisfy asymptotically the tail ratio
\begin{equation}
\lim_{x \rightarrow \infty} \frac{\overline{F}(x-y)}{\overline{F}(x)} = 1, \; \forall y \in [0,\infty).
\end{equation}
Severity models $F \in \mathcal{F}$ are of interest for severity distributions in high consequence loss modeling since they include models with \textit{infinite mean loss} and \textit{infinite variance}. In addition, the class $\mathcal{F}$ includes all severity models in which the tail distribution under the log transformed r.v., $\overline{F}\left(\log(x)\right)$, is a slowly varying function of $x$ at infinity.

To further understand LDA modeling with sub-exponential severity models we recall the notion of asymptotic equivalence in which a probability distribution function $F(x)$ is \textit{asymptotically equivalent} to another probability distribution function $G(x)$, denoted by $F(x) \sim G(x)$ as $x \rightarrow \infty$ if it holds that, $\forall \epsilon > 0, \exists x_0$ such that $\forall x > x_0$ the following is true
\begin{equation}
\left|\frac{F(x)}{G(x)} - 1\right| < \epsilon \; \text{ as } \; x \rightarrow \infty
\end{equation}
Furthermore, we say that a probability distribution function is \textit{max-sum-equivalent}, denoted by $F \sim_M G$, when the convolution of the tail distribution of two random variables is distributed according to the sum of the two tail distributions asymptotically, 
$$
1 - (F \star G)(x) = (\overline{F}\star \overline{G})(x) \sim \overline{F}(x) + \overline{G}(x), \;\; x \rightarrow \infty.
$$
Then for the class of heavy tailed sub-exponential LDA models we have that a probability distribution function $F$ will belong to the sub-exponential class $\mathcal{F}$ if $F \sim_M F$, i.e. it is max-sum-equivalent with itself and that the class $\mathcal{F}$ is closed under convolutions. The implications of this for the LDA models are clear when one observes that sub-exponential LDA models are compound process random sums comprised of an infinite mixture of convolved distributions,  
\begin{equation}
G(x) = \sum_{n=0}^{\infty} \lambda_n F^{n\star}(x),
\end{equation}
for a suitable series $\left\{\lambda_n\right\}$, (\textsl{e.g. convergent sequence satisfying Kolmogorov three series theorem}).

Proceeding in this section we will consider the compound Poisson distribution case in which the sequence $\lambda_n = e^{-\lambda}\frac{\lambda^n}{n!}$ and one can show the practically relevant asymptotic equivalence between the severity distribution $F$ and the annual loss distribution $G$ such that if $F \in \mathcal{F}$ then $G \in \mathcal{F}$ and 
$$
\lim_{x \rightarrow \infty} \frac{\overline{G}(x)}{\overline{F}(x)} = \lambda.
$$
These properties of sub-exponential LDA models regarding asymptotic equivalences make stochastic quantile and tail expectation approximations tractable for estimation. Only is special families of infinitely divisible severity models can closed form annual loss distributions be obtained, see discussions in \cite{peters2011impact} and \cite{peters2011analytic}. 

In general based on these properties we can obtain asymptotic approximations to the annual loss distribution tails which typically fall under one of the following classifications:
\begin{itemize}
\item{``First-Order'' and ``Second-Order'' Single Loss Approximations: recently discussed in \cite{bocker2009first}, \cite{degen2010calculation}, \cite{degen2011scaling} and references therein.}
\item{``Higher-Order'' Single Loss Approximations: see discussions in \cite{bingham1989regular} and recent summaries in \cite{albrecher2010higher} and references therein.}
\item{Extreme Value Theory (EVT) Single Loss Approximations (Penultimate Approximations): the EVT based asymptotic estimators for \underline{linear normalized} and \underline{power normalized} extreme value domains of attraction were recently discussed in \cite{degen2011scaling}.}
\item{Doubly Infinitely Divisible Tail Asymptotics given $\alpha$-stable severity models discussed in \cite{peters2011analytic} and \cite{peters2011impact}}
\end{itemize}

\textit{We now briefly detail the first and second order asymptotics known for light and heavy tailed severity distributions in LDA models, before then explaining how stochastic particle methods can be utilized to complement such closed form expressions and the role we believe will be played in the future by such stochastic algorithms in complementing these results.}

\subsubsection{The light tale of light tails}
At this stage it is very informative to understand the asymptotic results known for light tailed models as this will inform the results obtained in the heavy tailed expansions. An elegant summary of such a result was provided by \cite{embrechts1985asymptotic} where they consider frequency distributions $p_n = \mathbb{P}\mathrm{r}(N=n)$ satisfying 
\begin{equation*}
p_n \sim w^n n^{\gamma}C(n), \; n \rightarrow \infty.
\end{equation*}
for some $w \in (0,1)$, $\gamma \in \mathbb{R}$ and a function $C(n)$ slowly varying at $\infty$. If $\kappa > 0$ exists such that the Laplace transform of the severity
\begin{equation*}
L_{X}(s) = \mathcal{L}\left[F(x)\right] = \int_{0}^{\infty} \exp(-sx) dF(x), \; \forall s \in \mathbb{R},
\end{equation*}
matches the radius of convergence of the generating function of the frequency distribution,
\begin{equation*}
w^{-1} = L_{X}(-\kappa)
\end{equation*}
with $-L'_X(-\kappa) < \infty$, then the following asymptotic equivalence for the compound process tail distribution is satisfied,
\begin{equation*}
\overline{F}_{Z_N}(x) \sim \frac{x^{\gamma}\exp(-\kappa x)C(x)}{\kappa\left(-wL'_X(-\kappa)\right)^{\gamma + 1}}, \;\; x \rightarrow \infty.
\end{equation*}

This light tailed asymptotic result demonstrates that the behavior of the compound loss distribution tail is determined by either the frequency or the severity depending on which has the heavier tail. In addition it is clear that the Poisson distribution tail is too light for this result to be valid since the radius of convergence of generating function is infinite. There are therefore alternative expansions developed for compound Poisson risk processes such as the Saddle point approximation, see \cite{}. If the severity distribution is bounded, then as $x \rightarrow \infty$
\begin{equation*}
\overline{F}_{Z_N}(x) \sim \frac{\exp(\kappa x)}{|\kappa|\left(2\pi \lambda L''_{X}(\kappa)\right)^{1/2}}\left[\exp(-\lambda(1-L_X(\kappa)) - \exp(-\lambda)\right], 
\end{equation*}
where $\kappa$ is the solution of $- \lambda L'_{X}(\kappa) = x$.\\

\noindent \textbf{\textsl{So how do these results relate and motivate the context we are considering in sub-exponential LDA models}?}\\

Quite simply, in the sub-exponential heavy tailed setting the Laplace transform does not exist and hence these results do not apply. Examples of such models for which this is true include severity distributions with power law tail decay (\textsl{Pareto, Burr, log gamma, Cauchy, $\alpha$-Stable, tempered stable and t-distribution}).

\subsubsection{The heavier tale of heavy tails}
In this subsection we detail briefly the asymptotic fist and second order tail results for the LDA models when sub-exponential severity distributions are considered. The sub-exponential LDA first order tail asymptotics involve obtaining the closed form approximate expression for $\overline{F_{Z_N}}(x)$,
see details in \cite{bocker2005operational}, \cite{degen2011scaling}. To proceed, consider the annual loss distribution $G(z) = F_Z(z)$ under LDA formulation, given by,
\begin{equation*}
G(z) = F_Z(z) = \sum_{n=0}^{\infty} Pr\left[Z \leq z| N=n\right] Pr\left[N=n\right] = \sum_{n=0}^{\infty} p_n F^{(n)\star}(z),
\end{equation*}
with the severity distribution satisfying $F_X(z) \in \mathcal{F}$. Furthermore, assuming that for some $\epsilon > 0$,
\begin{equation*}
\sum_{n=0}^{\infty}\left(1+\epsilon\right)^n p_n < \infty.
\end{equation*}
Then the right tail of the annual loss distribution $F_Z(z)$ for the annual loss random variable $Z$, is approximated according to a SLA given by,
\begin{equation*}
\overline{F_Z}(x) = \mathbb{E}\left[N\right] \overline{F_X}(x) \left(1 + o(1)\right) \; \text{ as } \; x \rightarrow \infty,
\end{equation*}
or equivalently the tail of the annual loss distribution is sub-exponential in behavior with asymptotic equivalence, $$\overline{F_Z}(x) \sim \mathbb{E}[N] \overline{F_X}(x),$$.

To understand the basic result of the first order tail asymptotic $\overline{F_{Z_N}}(x)$ consider two steps:
\begin{enumerate}
\item{Obtain an upper bound on the asymptotic ratio of $\overline{F}_{Z_n}(x)$ and severity $\overline{F}(x)$ for all $n \in \mathbb{J}$. Typically one can apply Kesten's Bound which states that for subexponential severity distributions $F$ there exists a constant $K = K(\epsilon) < \infty$ for $\epsilon > 0$ s.t. $\forall n \geq 2$ the following bound holds (\cite{daley2007tail})
\begin{equation*}
\frac{\overline{F^{*n}}(x)}{\overline{F}(x)} \leq K(1+\epsilon)^n, \; x \geq 0.
\end{equation*}
} 
\item{Then simply utilize the Kesten bound to motivate the application of dominated convergence theorem to interchange the order of summation and limit and recall characterization of heavy-tailed sub-exponential severity models,
\begin{equation*}
\lim_{x \rightarrow \infty}\frac{\overline{F^{*2}}(x)}{\overline{F}(x)} = 2, \; \mathrm{implies} \; 
\lim_{x \rightarrow \infty}\frac{\overline{F^{*n}}(x)}{\overline{F}(x)} = n
\end{equation*}
}
\end{enumerate}
This process gives $\overline{F_{Z_N}}(x) \sim \mathbb{E}[N]\overline{F}(x)$ since:
\begin{equation*}
\lim_{x \rightarrow \infty} \frac{\overline{F_{Z_N}}(x)}{\overline{F}(x)} = \lim_{x \rightarrow \infty} \sum_{n=1}^{\infty} p_n \frac{\overline{F^{*n}}(x)}{\overline{F}(x)} = \sum_{n=1}^{\infty} n p_n =\mathbb{E}[N],
\end{equation*}

As discussed in \cite{degen2010calculation}, and the papers therein, the second order asymptotic results can be developed in a wide class of risk models by considering the following further assumptions.
\begin{enumerate}
\item[]{\textbf{Assumption 1: }\textsl{$F$ is zero at the origin ($x = 0$) and satisfies that both the tail distribution $\overline{F}$ and density $f$ are subexponential. }}
\item[]{\textbf{Assumption 2: }\textsl{The frequency distribution $N \sim F_N(n)$ is such that its probability generating function given by
\begin{equation*}
p_N(v) = \mathbb{E}\left[v^N\right] = \sum_{n=0}^{\infty} \mathbb{P}\mathrm{r}(N=n)v^n, 
\end{equation*}
is analytic at $v=1$}} 
\end{enumerate}
Examples of severity models that satisfy such assumptions include: Log-Normal, Weibull (heavy tailed), Benktander Type I and Type II, Inverse Gaussian, $\alpha$-Stable, Halphen Family, Normal-Inverse-Gaussian and other members of Generalized Hypergeometric family.

When distributions satisfy Assumption 1 and Assumption 2 then two situations may arise, those in which the loss random variable distribution is finite mean and the alternative case in which an infinite mean loss model is considered. If the loss r.v. has finite mean $(\mathbb{E}[X] < \infty)$ then the following result can be derived, see \cite{omey1986second} and \cite{willekens1989asymptotic} for details.
\begin{equation}
\lim_{x \rightarrow \infty} \frac{\overline{F}_Z(x) - \mathbb{E}[N]\overline{F}(x)}{f(x)} = \mathbb{E}[X]\mathbb{E}[(N-1)N].
\end{equation} 
Alternatively, if the loss r.v. is infinite but the severity density satisfies $f \in RV_{-1/\beta-1}$ for $1 \leq \beta < \infty$ then:
\begin{equation*}
\lim_{x \rightarrow \infty} \frac{\overline{F}_Z(x) - \mathbb{E}[N]\overline{F}(x)}{f(x)\int_{0}^x \overline{F}(s)ds} = c_{\beta}\mathbb{E}[(N-1)N].
\end{equation*}
with $c_{1} = 1$ and $c_{\beta} = (1-\beta) \frac{\Gamma^2(1-1/\beta)}{2\Gamma(1-2/\beta)}$ for $\beta \in (1,\infty)$. We recall that a measurable function $f: (0,\infty) \rightarrow (0, \infty)$ is called regularly varying (at infinity) with index $\beta$ and denoted by $f \in RV_{-1/\beta-1}$ if $\forall u > 0$, $\exists \beta \in \mathbb{R}$ where
\begin{equation*}
\lim_{x \to \infty} \frac{f(ux)}{f(x)}=u^{\beta}.
\end{equation*}

In the following subsection we clearly detail how and why such asymptotic results are utilized for inference in risk and insurance, before highlighting the important potential role stochastic particle methods will play in complementing these results.

\subsection{Inferential Challenges for Risk and Insurance:\\ Asymptotics and the Role for Stochastic Particle Integration}
The asymptotic approximation methods just surveyed were developed in the actuarial literature to tackle the serious statistical and computational challenges posed by estimation of tail quantiles and expectations for heavy-tailed LDA models. The continued interest in such asymptotic results primarily stems from the fact that such closed form expressions bypass the serious computational challenges for estimation of risk measures for such heavy-tailed annual loss distributions under traditional integration methods, Fourier methods, recursions (Panjer) or basic Monte Carlo approaches. However, they do have associated issues, see discussions in \cite{hess2011can}. 

The properties of such asymptotic single loss approximation estimates are still an active subject of study with regard to aspects such as explicit approximation errors, unbiased quantile function estimation, asymptotic rates of convergence, sensitivity to parameter estimation and model misspecification. It is in the understanding of these increasingly important practical features that we believe stochastic particle integration methods will complement the asymptotic results, which are generally un-attainable under such additional considerations.

Before introducing in depth some key results in stochastic particle methods, we will first complete the risk and insurance motivation by explaining the key result one obtains from a risk management perspective as a consequence of these first and second order asymtptotics. We will also demonstrate exactly how it is often utilized to make risk management based decisions by tying it back to the calculation of risk measures in LDA models. 

Based on the results obtained for the second order asymptotic in the heavy tailed LDA models, one can show that if the severity distribution $F$ satisfies Assumption 1 and Assumption 2 with a finite mean, and the hazard rate $h(x) = \frac{f(x)}{1-F(x)}$ is of regular variation $h \in RV_{-\beta}$ for $\beta \geq 0$, then as $\alpha \rightarrow 1$ one has for the inverse of the annual loss distribution the result
\begin{equation}
F^{-1}_{Z}(\alpha) = F^{-1}\left(1 - \frac{1 - \alpha}{\mathbb{E}[N]}\left\{1 + \widetilde{c}_{\beta}g_1\left(F^{-1}(\widetilde{\alpha})\right) + o\left(g_1\left(F^{-1}(\widetilde{\alpha})\right) \right) \right\}^{-1} \right)
\end{equation}
where $\widetilde{\alpha} = 1 - (1-\alpha)/\mathbb{E}[N]$ and
\begin{equation*}
\begin{split}
g_1(x) &= \begin{cases}
\frac{f(x)}{1-F(x)}, & \mathrm{if} \, \mathbb{E}[X] < \infty,\\
\frac{\int_{0}^x\overline{F}(s)ds f(x)}{1 - F(x)}, & \mathrm{if} \, \mathbb{E}[X] = \infty.;
\end{cases}\\
\widetilde{c}_{\beta} &= \begin{cases}
\frac{\mathbb{E}[X]\mathbb{E}[(N-1)N]}{\mathbb{E}[N]}, & \mathrm{if} \, \mathbb{E}[N] < \infty,\\
\frac{c_{\beta}\mathbb{E}[(N-1)N]}{\mathbb{E}[N]}, & \mathrm{if} \, \mathbb{E}[N] = \infty.\\
\end{cases}
\end{split}
\end{equation*}
From this result it is then possible to consider asymptotic approximations of key risk management quantities known as risk measures which are used in the allocation of capital and reserving in all financial institutions and stipulated as standards under regulatory accords in both Basel II/III and Solvency II. Examples of such tail functionals include the calculation of Value-at-Risk (VaR), Expected Shortfall (ES) and Spectral Risk Measures as detailed below in both their definitions and the resulting simple asymptotic approximations one may consider. 

These asymptotic expansions allow one to obtain estimates of common risk measures, see \cite{artzner1999coherent} and \cite{mcneil2005quantitative}, such as
\textit{Value-at-Risk (VaR)} for a level $\alpha \in (0,1)$, given by the quantile of the annual loss distribution,
\begin{equation}
\begin{split}
\mathrm{VaR}_{Z}\left(\alpha\right) &= F^{\leftarrow}_{Z}(\alpha) = \inf\left\{z \in \mathbb{R}:F_{Z}(z) \geq \alpha \right\}\\
&\approx F_Z^{\leftarrow}\left(1-\frac{1-\alpha}{\mathbb{E}[N]}\left[1 + o(1) \right]\right) \approx F^{\leftarrow}\left(1-\frac{1-\alpha}{\mathbb{E}[N]}\right),
\end{split}
\end{equation}
where $F^{\leftarrow}(\cdot)$ is the generalized inverse, see \cite{embrechts2010note}. The \textit{Expected Shortfall (ES)}, see \cite{biagini2009asymptotics}, for a level $\alpha \in (0,1)$ is given by the tail expectation of the annual loss distribution according to
\begin{equation}
\begin{split}
\mathrm{ES}_{Z}(\alpha) &= \mathbb{E}\left[Z|Z \geq \mathrm{VaR}_{Z}\left(\alpha\right)\right] = \frac{1}{1-\alpha}\int_{\alpha}^1 \mathrm{VaR}_Z(s)ds\\
&\approx \frac{\alpha}{\alpha - 1}F^{\leftarrow}\left(1-\frac{1-\alpha}{\mathbb{E}[N]}\right) \sim \frac{\alpha}{\alpha - 1}VaR_{Z}\left(\alpha\right),
\end{split}
\end{equation}
and the \textit{Spectral Risk Measure (SRM)} for a weight function $\phi:[0,1] \mapsto \mathbb{R}$ given by 
\begin{equation}
\begin{split}
\mathrm{SRM}_{Z}(\phi) &= \int_{0}^1 \phi(s) \mathrm{VaR}_Z(s)ds\\
&\approx \mathcal{K}(\alpha,\phi_1)F^{\leftarrow}\left(1-\frac{1-\alpha}{\mathbb{E}[N]}\right) \sim  \mathcal{K}(\alpha,\phi_1)VaR_{Z}\left(\alpha\right),
\end{split}
\end{equation}
with $\forall t \in (1,\infty)$ a function $\phi_1(1-1/t) \leq K t^{-1/\beta + 1 - \epsilon}$ for some $K>0$ and $\epsilon > 0$ where
$$
\mathcal{K}(\alpha,\phi_1) = \int_{1}^{\infty} s^{1/\beta - 2}\phi_1(1-1/s)ds.
$$

\subsubsection{The Role for Stochastic Particle Methods}
Though the asymptotic results presented are elegant and efficient to evaluate, they do warrant careful consideration in their application. In this section we explain what we mean by this statement and then utilize this to motivate the use of stochastic particle methods. It is important for practitioners to understand that the properties of such SLA estimates is still a subject of study, this includes for example an understanding of (approximation error, unbiased quantile function estimation, asymptotic rates of convergence, sensitivity to parameter estimation, model misspecification etc.) all of which can have a non-trivial influence on the resulting asymptotics and therefore risk measure estimates, as discussed recently in \cite{EmbrechtModel}.

In practice it may often be the case that one requires calculation of VaR, ES and Spectral Risk Measures at levels which do not satisfy such asymptotic properties, rendering such approximations inaccurate. In addition, though not yet a regulatory requirement, it is always good practice to consider the uncertainty associated with the estimation of the tail functionals and quantiles, through say confidence intervals and this is non-trivial to obtain under such asymptotic expansion results. Thirdly, as discussed in \cite{albrecher2010higher} and \cite{barbe2009asymptotic} the asymptotic rates of convergence of such approximations are still only known in a little-oh Landau sense and therefore do not inform or guide the applicability of such results. There is also a significant complication with such asymptotic results that arises in the application of such results for models that in practice do not admit closed form representations of the quantile function of the severity distribution in the LDA model. Finally, there is also a significant interest in diversification benefits that may be gained through the modeling of tail dependence features in the multi-variate risk process setting. Whilst these asymptotic results are extendable to the multi-variate case of multiple risk processes, the addition of even parametric tail dependence through a copula renders the derivation of such results highly challenging and an active research area at present, see for example recent results in \cite{hua2012tail}.

It is in these four key elements that we argue stochastic particle based numerical solutions to such inference on risk measures and tail functionals can be of direct utility to complement such asymptotic results. \textsl{However, as all practitioners will know, the naive implementation of standard Monte Carlo and stochastic integration approaches to such problems will produce often poor results even for a considerable computational budget, see discussions in \cite{luo2009computing}.} There is therefore a computational challenge for estimation of risk measures for such heavy-tailed annual loss distributions that we argue can be addressed by stochastic particle methods. This concludes the discussion on motivations for how such methods can play an increasingly more important role in risk and insurance and we now present a detailed exposition of a few important stochastic particle methods that will be a solid starting point for practitioners.

\section{Selected Topics in Stochastic Integration Methods}\label{sec:2}
In this section we will introduce a variety of stochastic integration methods, presenting them formally from a mathematical perspective and making clear the properties of such methods. This will provide practitioners with an understanding of the key properties of these methods and the relevant references to consider in applying such approaches to tackling risk and insurance problems. Note, in this section the notation adopted is utilized to reflect that which is considered in the statistics and probability literature where much of the formal study of these methods has taken place. We first introduce examples of problem domains where each approach has been considered in the risk and insurance literature, to tackle particular inference problems for specific models, before formally detailing the general mathematical understanding of these methods.

\subsection{Standard Monte Carlo Techniques for Risk and Insurance}
Let $X$ be a $d$-dimensional random variable and $A\subset\RR^d$ some measurable subset. Suppose we want to compute the quantity $\PP(X\in A):=\PP_X(A)$. We further assume that it is straightforward to generate a sequence  $(X^i)_{1\leq i\leq N}$ of independent copies of the random variable $X$. In this situation, the traditional Monte Carlo approximation of the distribution $\PP_X$ is given by the empirical measures
$$
\PP_X^N=\frac{1}{N}\sum_{1\leq i\leq N}\delta_{X^i}\longrightarrow_{N\uparrow\infty}  \PP_X
$$
More precisely, the convergence can be understood as the weak convergence of empirical measures, in the sense that the following convergence holds
$$
\PP_X^N(f):=\int f(x)~ \PP_X^N(dx)=\frac{1}{N}\sum_{1\leq i\leq N}f(X^i)\longrightarrow_{N\uparrow\infty}  \PP_X(f)=\int f(x)~ \PP_X(dx)=\EE(f(X))
$$
almost surely, for any bounded measurable function $f$ on $\RR^d$. Using indicator functions of cells in $\RR^d$, the shape of the measure $\PP_X$ can be obtained by plotting the histograms of the samples $X^i$ in every dimensions. By the strong law of large numbers, the above convergence is also met for integrable functions w.r.t. the measure $\PP_X$.  
 
For indicator functions $f=1_A$, sometimes we make a slight abuse of notation and we set $ \PP_X^N(A)$ and $ \PP_X(A)$ instead of  $ \PP_X^N(1_A)$ and $ \PP_X(1_A)$. From the above discussion, we already have that
$$
 \PP_X^N(A):=\frac{1}{N}\sum_{1\leq i\leq N}1_{A}(X^i)~\longrightarrow_{N\uparrow\infty}  \PP_X(A)=\EE(1_A(X)).
$$
The following properties are readily checked
 $$
 \EE( \PP_X^N(A))= \PP_X(A)
 \quad\mbox{\rm and}\quad
\mbox{\rm Var}\left( \PP_X^N(A)\right)=\frac{1}{N}~\PP_X(A)~\left(1-\PP_X(A)\right).
$$
 In addition,  an $N$-approximation of the conditional distribution of $X$ w.r.t. the event 
 $\{X\in A\}$ is given by
\begin{equation}\label{MC1}
\frac{1}{\PP_X^N(A)}~1_A(x)~\PP_X^N(dx)\longrightarrow_{N\uparrow\infty}\frac{1}{\PP_X(A)}~1_A(x)~\PP_X(dx)=\PP\left(X\in dx~|~X\in A\right).
\end{equation}

The l.h.s. terms in the above display is well defined as soon as $\PP_X^N(A)>0$.  For rare event probabilities $\PP_X(A)$, say of order $10^{-6}$, the practical implementation of this Monte Carlo algorithm meets the difficulty that we need too many samples to estimate $\PP_X(A)$  using the proportion of success of such an event occurring only once per millions of attempts. 

We illustrate this on a standard model in risk and insurance based on the Poisson-Log Normal LDA model of a single risk process. This example though simple is both widely utilized in practice and also illustrative of the complementary role of the asymptotic approximations and the role Monte Carlo plays, since this specific model admits a closed form expression for the survival quantile of the annual loss under the first order asymptotic. 

\begin{example}[Single Risk LDA Poisson-Log-Normal Family]{
Consider the heavy-tailed severity model, selected to model the sequence of i.i.d. losses in each year $t$, denoted $\left\{X_i(t)\right\}_{i=1:N_t}$, and chosen to be a Log-Normal distribution $X_i \sim LN(\mu, \sigma)$ where the two parameters in this model correspond to parameterizing the shape of the distribution for the severity $\sigma$ and the log-scale of the distribution $\mu$. The survival and quantile functions of the severity are given by
\begin{equation*}
\begin{split}
&f_X(x;\mu,\sigma) = \frac{1}{x\sqrt{2\pi\sigma^2}}\, e^{-\frac{\left(\ln x-\mu\right)^2}{2\sigma^2}} , \; x>0; \; \mu \in \mathbb{R} \; \sigma >0 \\
&\bar{F}(x;\mu,\sigma) = 1-F(x)= \int_{x}^{\infty}\frac{1}{\sqrt{2 \pi \sigma u}}\exp\left( -\frac{1}{2 \sigma^2}\left( \log(u) - \mu^2\right) \right) du\\
& \hspace{1.6cm} =\frac12 + \frac12\,\mathrm{erf}\Big[\frac{\ln x-\mu}{\sqrt{2\sigma^2}}\Big] , \; x>0; \; \mu \in \mathbb{R} \; \sigma >0 \\
&Q(p) = \exp\left(\mu + \sigma \Phi^{-1}(p) \right), \; 0<p<1.
\end{split}
\end{equation*}
Therefore the closed form SLA for the VaR risk measure at level $\alpha$ would be presented in this case under a first order approximation for the annual loss $Z = \sum_{n=1}^N X_i$ according too Equation \ref{Eqn:VaRLNSLA}
\begin{equation} \label{Eqn:VaRLNSLA}
\mathrm{VaR}_{\alpha}\left[Z\right] = \exp\left[\mu - \sigma \Phi^{-1}\left(\frac{1-\alpha}{\lambda}\right) \right]
\end{equation}
To compare this first order asymptotic result to the crude Monte Carlo approach (for which one can generate uncertainty measures such as confidence intervals in the point estimator) it is first required to detail how to simulate such an annual loss process. In this simple example the simulation of a loss process from a Log-Normal severity distribution can be achieved via a transformation of a Gaussian random variate, which itself is generated typically via a transformation of two uniform random variates in a Box-Muller approach, see details in \cite{devroye1986non}. The basic Monte Carlo simulation of the annual loss process for $T$-years, from a Poisson-Log-Normal LDA model can be achieved via a transformation of standard random variates as follows:\\
\\
\noindent \textbf{Algorithm 1: Poisson-Log Normal LDA model via Standard Monte Carlo}
\begin{enumerate}
\item{Generate vector of realized annual loss counts $\mathbf{N}_{1:T} = \left\{N_1,N_2,\ldots,N_T\right\}$ by drawing from a Poisson distribution with rate $\lambda$, $N_t \sim Po(\lambda)$. This is undertaken for each random variate realization $N_t = n_t$ for each year of simulation $t \in \left\{1,2,\ldots,T\right\}$ via one of several possible algorithms, such as:}
\begin{enumerate}
\item{Set $L = \exp(-\lambda), s=0$ and $p=1$.}
\item{While $p > L$: set $s = s + 1$ then generate $U \sim U[0,1]$ and set $p = pU$.}
\item{Set $n_t = s$}
\end{enumerate}
\item{For each $t$, generate $n_t$ i.i.d realizations of the loss $X_i(t) \sim LN(\mu,\sigma)$ via transformation according to:}
\begin{enumerate}
\item{Generate two independent uniform random numbers $U_1$, $U_2$ uniformly distributed on $(0, 1]$} 
\item{Apply the Box-Muller transformation
\begin{equation}
Y = \mu + \sigma\left(\sqrt{12 \ln\left(U_2\right) } \cos\left(2 \pi U_1\right) \right) \sim N\left(\mu,\sigma^2\right). 
\end{equation}
}
\item{Apply the transformation from Normal to Log-Normal
\begin{equation}
X = \exp(Y) \sim LN\left(\mu,\sigma\right). 
\end{equation}
}
\end{enumerate}
\item{Set the realized annual loss in each year $t$ to $Z(t) = \sum_{i=1}^{N_t}X_i(t)$.}
\end{enumerate}
To complete this example, we illustrate the basic Monte Carlo solution for the VaR for a range of quantile levels of the annual loss distribution along with the measures confidence intervals in the point estimators,  compared to the first order asymptotic result. The quantiles $\alpha \in \left\{0.70, 0.75, 0.80, 0.85, 0.9, 0.95, 0.99, 0.995, 0.9995\right\}$ are considered where the $99.5\%$ and $99.95\%$ quantile levels do in fact correspond to regulatory standards of reporting in Basel II/III.
\begin{figure}[h]
\label{RHIE_Chpt:LN}
\includegraphics[height=10cm, width=\textwidth] {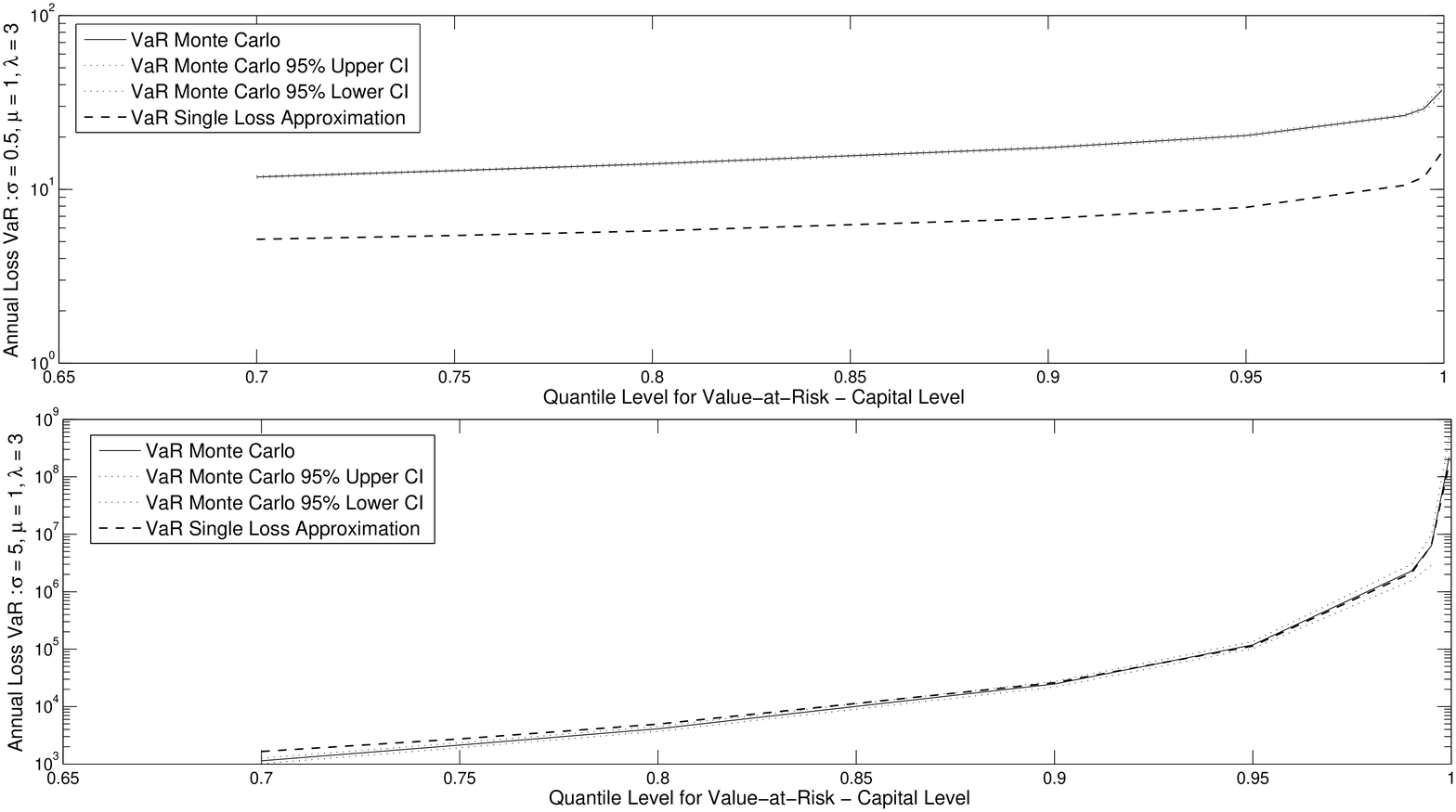}
\caption{Annual Loss VaR Capital Estimate versus quantile level for Poisson-Log Normal LDA Risk Process.
\textbf{Top Plot:} Severity distribution $\mu = 1, \sigma = 0.5, \lambda = 3$.
\textbf{Bottom Plot:} Severity distribution $\mu = 1, \sigma = 5, \lambda = 3$.}
\end{figure}
}\end{example}

This example illustrates clearly the motivation for consideration of more developed methods, since one can see that even in this relatively simple example, depending on the values of the parameters in the LDA risk model, the asymptotic VaR approximation may or may not be accurate at quantile levels of interest to risk management. Clearly in the top sub-plot the accuracy of a capital estimate formed from the SLA will be poor where as the accuracy from the second subplot with the different risk profile parameters is fine for quantiles beyond the 80-th percentile. However, in general since the rate of convergence is still an active topic of research for such approximations, the only way to ensure accuracy of such methods for a given set of estimated / specified parameters in practice is to complement these approximations with a numerical solution or comparison. In general such an approach will require more sophisticated numerical procedures for more challenging risk process settings. In this example we utilized 1,000,000 annual years and the Mote Carlo accuracy was sufficient, note as $\lambda$ decreases, as will be the case for high consequence loss events that one considers modeling with sub-exponential models, this number of samples will need to significantly increase.

\subsection{Importance Sampling Techniques for Risk and Insurance}
One could argue that the second most widely utilized class of stochastic integration methods utilized in risk and insurance settings would have to be the Importance Sampling family, see for example reviews with control variates in market risk settings in \cite{glasserman2003Monte}. In the insurance setting one can see for example \cite{peters2007simulation}.

To understand these class of methods as another way to design a feasible algorithm is to sample using another random variable for which the occurrence probability of the desired event $\PP(Y\in A):=\PP_Y(A)$ is closer to $1$. This well known importance sampling strategy often gives efficient results for judicious choices of twisted measures $\PP_Y$. Nevertheless, in some practical situations, it is impossible to find a judicious $\PP_Y$ that achieves a given efficiency. Furthermore, this importance sampling technique is intrusive, in the sense that it requires to change the reference statistical or physical model into a twisted sampling rule. 

To be more precise, sampling $N$ independent copies 
 $(Y^i)_{1\leq i\leq N}$ with the same dominating probability measure $\PP_Y\gg\PP_X$,  the traditional Monte Carlo approximation is now 
 given by
$$
 \PP_Y^N\left(1_A~\frac{d\PP_X}{d\PP_Y}\right):=\frac{1}{N}\sum_{1\leq i\leq N}~1_{A}(X^i)~\frac{d\PP_X}{d\PP_Y}(Y^i)~\longrightarrow_{N\uparrow\infty}
\PP_Y\left(1_A~\frac{d\PP_X}{d\PP_Y}\right)= \PP_X(A)
$$
The following properties are readily checked
 $$
 \EE\left(  \PP_Y^N\left(1_A~\frac{d\PP_X}{d\PP_Y}\right)\right)= \PP_X(A)
$$
and
$$
\mbox{\rm Var}\left(  \PP_Y^N\left(1_A~\frac{d\PP_X}{d\PP_Y}\right)\right)=\frac{1}{N}~\left(
\PP_X\left(1_A~\frac{d\PP_X}{d\PP_Y}\right)-\PP_X(A)^2
\right)
$$
It is easily check that
$$
\PP_Y(dx)=\frac{1}{\PP_X(A)}~1_A(x)~\PP_X(dx)\Rightarrow\mbox{\rm Var}\left(  \PP_Y^N\left(1_A~\frac{d\PP_X}{d\PP_Y}\right)\right)=0
$$
In other words, the optimal twisted measure $\PP_Y$ is given by the unknown conditional distribution of $X$ w.r.t. the event 
 $\{X\in A\}$. In practice, we try to find a judicious choice of twisted measure that are easy to sample, with a probability mass distribution
 that resembles as much as
 possible to the desired conditional distribution.
 
Another traditional idea is to
 use the occupation measure of a judiciously chosen Markov Chain Monte Carlo ({\em abbreviate MCMC}) sampler with prescribed target measure 
 $$
\eta(dx):=\PP\left(X\in dx~|~X\in A\right).
 $$
 
Of course,  the first candidate is to take a sequence of independent copies of random variables with common distribution $\eta$. Several exact sampling techniques can be used, including the inversion of the repartition function, change of variables principles, 
the coupling from the past, and acceptance-rejection techniques.  
For instance, the Monte Carlo approximation presented in (\ref{MC1}) is clearly 
based on this universal and traditional acceptance-rejection sampling technique. A random sample $X_i$ with distribution $\PP_X$ is accepted whenever it enters in the desired subset $A$. In this interpretation,  we need to sample
$N$ independent copies of $X$ to obtain
$\overline{N}:=N\times \PP_X^N(A)$ independent samples with common law $\eta$. However, for probabilities $\PP_X(A)$
of order $10^{-6}$, this method requires millions of samples. 

\subsection{Markov chain Monte Carlo for Risk and Insurance}
After the class of standard Monte Carlo and the Importance Sampling techniques one would naturally consider the next most widely developed class of methods as the Markov Chain Monte Carlo methods. These have found use in insurance applications in non-life reserving models for example in Chain Ladder models \cite{peters2009model}, \cite{peters2010chain}, \cite{england2010bayesian} and Paid Incurred Claims models \cite{merz2010paid} and \cite{peters2012copula}, in Operational Risk models in \cite{peters2009dynamic}, \cite{peters2006bayesian} and in credit risk modeling for example in \cite{luo2010lgd}.

Hence, we now present the fundamental mathematical description of the underlying Monte Carlo algorithm that is developed for all the risk and insurance applications above for specific models. More generally, MCMC algorithms are based on sampling a Markov chain with invariant measure $\eta$. In this context, the limiting measure $\eta$ is often called the target measure. It is not difficult to construct these random processes.
For instance, let us assume that the law of $X$ is reversible w.r.t. some Markov transition $K(x,dy)$. In this case, starting from the set $A$, we sample a sequence of random states using the Markov proposal $K$, rejecting sequentially all the states falling outside the set $A$. The algorithm is well defined as soon as $K(x,A)=K(1_A)(x)>0$, and the resulting Markov chain $X_{n}$ coincides with the Metropolis-Hasting algorithm with probability transition given by the following formulae
$$
M(x,dy):=K(x,dy)~1_A(y)+\left(1-\int~K(x,dz)~1_A(z) \right)~\delta_{x}(dy).
$$

It is not difficult to check that $\eta$ is an invariant measure of the chain with transition $M$, that is we have that
$$
(\eta M)(dy):=\int \eta(dx)~M(x,dy)=\eta(dy).
$$

The  exact acceptance-rejection method discussed above and in (\ref{MC1}) corresponds to the special case 
$$
K(x,dy)=\PP\left(X\in dy\right)
$$

In more general situations, the proposal transition $K(x,dy)$ amounts of moving randomly around the starting point $x$.
The individual (sometimes also called the walker) makes a number of tentative steps until it succeeds to enter into the desired set
$A$. In general, the random state at that (random) hitting time of $A$ is not distributed according to $\eta$. Roughly speaking, when the proposal
transition $K$ is based on local moves, the individual tends to hit the set $A$ near the boundary of A. To be more precise, starting from an initial state $X_0=x\in\RR^d-A $ 
the hitting time 
$$
T:=\inf{\left\{n\geq 0~:~X_n\in A \right\}}
$$ 
is a geometric random variable with distribution
$$
\PP\left(T=n~|~X_0=x\right)=\left(1-K(x,A)\right)^{n-1}~K(x,A)
$$
and we have
$$
\EE\left(f(X_T)~|~X_0=x\right)=K_A(f)(x):=K(f1_A)(x)/K(1_A)(x).
$$

When the chain enters in $A$, it remains for all times confined to the set $A$. 
In  addition, under some weak regularity conditions on the Markov transition $K$, 
the target measure $\eta$ is approximated by the occupation measures of the states; that is, we have the following asymptotic 
convergence result
\begin{equation}\label{theo-ergo}
\frac{1}{n+1}\sum_{0\leq p\leq n}\delta_{X_p}\longrightarrow_{n\uparrow\infty}  \eta\quad\mbox{\rm and}\quad
\PP\left(X_n\in dy~|~X_0=x \right):=M^{n}(x,dy)\longrightarrow_{n\uparrow\infty}  \eta(dy).
\end{equation}
In the above display, $M^{n}(x,dy)$ stands for the $n$ compositions of the integral operator $M$ defined by the induction
formulae
$$
M^{n}(x,dy)=\int M^{n-1}(x,dz)M(z,dy)=\int M(x,dz)M^{n-1}(z,dy)
$$
with the convention $M^{0}(x,dy)=\delta_x(dy)$, for $n=0$.

It is of course out of the scope of this article to prove the ergodic theorem stated in the l.h.s. of (\ref{theo-ergo}). We end this section
with a simple proof of the r.h.s. assertion. Firstly, we observe that
\begin{eqnarray*}
M^n(f)(x)&=&\EE\left(f(X_n)~1_{T<n}~|~X_0=x\right)+\EE\left(f(X_n)~1_{T\geq n}~|~X_0=x\right)\\
&=&\sum_{1\leq k<n}(1-K(x,A))^{k-1}K(x,A)~K_A(M^{n-k}(f))(x)\\
&&\hskip3cm+f(x)\sum_{k\geq n}~(1-K(x,A))^{k-1}K(x,A).
\end{eqnarray*}
On the other hand, we have
 \begin{eqnarray*}
 1_A(x) M(x,dy)=1_A(x) M(x,dy)1_A(y)&\Longrightarrow &1_A(x)~M(1_A)(x)=1_A(x)\\
 &\Longrightarrow& 1_A(x)M(x,dy)=1_A(x) M_A(x,dy)
\end{eqnarray*}
 with the Markov transitions
 $$
 M_A(x,dy)=\frac{M(x,dy)1_A(y)}{M(1_A)(x)}~~\left(=K_A(x,dy)~~\mbox{\rm if }~~ x\not\in A\right).
 $$
This clearly implies that
$$
1_A M^{m}(f)=1_A M_A^{m}(f)\Rightarrow K_AM^{m}=K_AM_A^m\quad\mbox{\rm and}\quad
\eta M=\eta M_A^m
$$
from which we find that
$$
K_AM^{m}(f)(x)-\eta(f)=\int~K_A(x,dy)\eta(dy^{\prime})~\left[M^m_A(f)(y)-M^m_A(f)(y^{\prime})\right].
$$
This implies that
$$
\sup_{x\in\RR^d}\left\|K_AM^{m}(x,\point)-\eta\right\|_{\tiny tv}\leq \beta(M^m_A):=
\sup_{y,y^{\prime}\in A}{\left\|M^m_A(y,\point)-M^m_A(y^{\prime},\point)\right\|_{\tiny tv}}.
$$
In the above display, $\| \mu_1-  \mu_2\|_{\tiny tv}$ stands for the total variation distance between two probability measures
$\mu_1$, and  $\mu_2$ defined by
$$
\| \mu_1-  \mu_2\|_{\tiny tv}
=\sup{\left\{[\mu_1-  \mu_2](f)~:~\mbox{\rm osc}(f)\leq 1 \right\}}.
$$

We consider the following mixing condition

$(H_A)$ There exists some  probability measure $\nu$ on $\RR^d$, and some $ \epsilon_A\in]0,1]$ such that
$$
\forall x\in A\quad M_A(x,dy)\geq \epsilon_A~\nu(dy)
$$

This condition is clearly met as soon as
$$
\forall x\in A\quad K(x,dy)~1_A(y)\geq \epsilon_A~\nu(dy)
$$
For instance, when $K(x,dy)$ has a density $k(x,y)$ w.r.t. the Lebesgue measure $\lambda(dy)$ on $\RR^d$, the condition
 $(H_A)$ is met as soon as $\overline{k}(y):=\inf_{x\in A}k(x,y)$ is s.t. $\lambda(\overline{k})>0$. In this case,  $(H_A)$ is met with
 $\epsilon_A=\lambda(\overline{k})$ and $\nu(dy)\propto\overline{k}(y)~\lambda(dy)$.

We also recall that
\begin{equation}\label{mixing}
(H_A) \Longrightarrow \beta(M_A^m)\leq (1-\epsilon_A)^m.
\end{equation}
Next, we provide a short proof  (\ref{mixing}). Under the r.h.s. condition, the following Markov transition
$$
M^{\prime}_A(x,dy):=\frac{ M_A(y,dx)- \epsilon_A~\nu(dx)}{1-\epsilon_A}
$$
is well defined, and we have
$$
M_A(f)(y)-M_A(f)(y^{\prime})=\left(1-\epsilon_A\right)\left(M^{\prime}_A(f)(y)-M^{\prime}_A(f)(y^{\prime})\right)\Rightarrow
\beta(M_A)\leq (1-\epsilon).
$$

Iterating the argument, we readily prove that for any $m\geq 1$
$$
\sup_{y,y^{\prime}\in A}{\left\|M^m_A(y,\point)-M^m_A(y^{\prime},\point)\right\|_{\tiny tv}}\leq (1-\epsilon_A)^m.
$$
Using the decomposition
\begin{eqnarray*}
M^n(f)(x)-\eta(f)
&=&\sum_{1\leq k<n}(1-K(x,A))^{k-1}K(x,A)~\left(K_A(M^{n-k}(f))(x)-\eta(f)\right)\\
&&\hskip3cm+\left(f(x)-\eta(f)\right)\sum_{k\geq n}~(1-K(x,A))^{k-1}K(x,A)
\end{eqnarray*}
we prove that
\begin{eqnarray*}
\|M^n(x,\point)-\eta\|_{\tiny tv}&\leq &\sum_{1\leq k<n}(1-K(x,A))^{k-1}K(x,A)(1-\epsilon_A)^{n-k}\\
&&\hskip3cm+\sum_{k\geq n}~(1-K(x,A))^{k-1}K(x,A).
\end{eqnarray*}
After some elementary computations, we conclude that
$
\lim_{n\uparrow\infty}\|M^n(x,\point)-\eta\|_{\tiny tv}=0
$.

\subsection{Sequential Monte Carlo for Risk and Insurance} \label{Section:SMCMethods}
The application of Sequential Monte Carlo (SMC) methods in the study of important problems in risk and insurance modeling is still relatively under developed, hence the motivation for this article. In the context of risk modeling see the example in \cite{peters2009sequential} and the references therein for more discussion. We start this section with a motivating class of algorithms targeting rare-event simulation via the restriction of a target measure to a contracting, increasingly rare set, such as a tail event. We detail the formal mathematical description of such a particle system stochastic algorithm, which is often omitted in the application papers utilizing such methods and some times to the detriment of both understanding and validity of results obtained. We believe it is important to present the principled understanding of this class of numerical integration regimes to clearly understand their utility in solving inferential problems in risk and insurance settings.

\textsl{However, we note that whilst this is a good motivation for these methods in risk and insurance, it is by no means the only method of SMC in risk and insurance applications as we illustrate in the class of algorithm developed in Section \ref{SectionExample}.}

Sequential Monte Carlo methods are acceptance-rejection techniques equipped with a recycling mechanism
that allows to sample gradually a population of individuals w.r.t. a sequence of probabilities with increasing complexity. We illustrate this methodology in the situation discussed above. Let us
 choose a decreasing sequence of subsets $(A_p)_{0\leq p\leq n}$ joining  
 $A_0=\RR^d$ to the desired lower subset $A_n=A$:
$$
A_0=\RR^d\subset A_1\subset A_2\subset\ldots\subset A_{n-1}\subset A_n=A.
$$ 
Now, let's try to sample sequentially random copies of  the random variable $X$ w.r.t the conditioning events 
$\{X\in A_p\}$, with $p\leq n$.  To get one step further, we let $\eta_p$ be the sequence
of measures 
$$
\eta_p(dy):=\PP\left(X\in dx~|~X\in A_p\right)\quad\mbox{\rm with}\quad p\leq n.
$$
By construction, $\left(\eta_p\right)_{0\leq p\leq n}$ is a decreasing sequence of measures w.r.t. the 
absolutely continuous partial order relation $\mu\ll\nu$ between probability 
measures  \footnote{we recall that $\mu\ll\nu$ as soon as $\nu(A)=0\Rightarrow \mu(A)=0$, for all measurable subset $A\subset \RR^d$}; that is, we have that 
$$
\eta_n\ll\eta_{n-1}\ll\ldots\ll
\eta_2\ll\eta_1\ll\eta_0=\mbox{\rm Law}(X).
$$

\subsubsection{Sequential Markov chain Monte Carlo methods}
In this connection, we further assume that we have a dedicated MCMC style probability 
 transitions $M_p$ with invariant measure $\eta_p=\eta_p M_p$, for any $p\leq n$. 
We start running a sequence of random states $(X_p)_{0\leq p\leq n_1}$ with transitions $M_1$, and initial condition $\eta_0$.
For a sufficiently large
 time horizon $n_1$, both the occupation measure $\frac{1}{n_1}\sum_{1\leq p\leq n_1}\delta_{X_p}$
 and the law of the terminal state $\mbox{\rm Law}(X_{n_1})=\eta_0 M_1^{n_1}:=\pi_1$  approximate
 the target measure $\eta_1$.  
We also notice that the chain $(X_p)_{p_1\leq p\leq n_1}$ is confined to the set $A_1$ as soon as one of the random states $X_{p_1}\in A_1$ hits the set 
$A_1$ for some $p_1\leq n_1$.

 In the second step, starting from  $X_{n_1}$ we run a sequence of random states $(X_{n_1+p})_{0\leq p\leq n_2}$ with transitions $M_2$ (and initial condition $\pi_1$).  
 For a sufficiently large
 time horizon $n_2$, both the occupation measure $\frac{1}{n_2}\sum_{1\leq p\leq n_1}\delta_{X_{n_1+p}}$
 and the law of the terminal state $\mbox{\rm Law}(X_{n_1+n_2})=\pi_1 M_2^{n_2}$  approximate
 the target measure $\eta_2$.  As before, the chain $(X_{n_1+p})_{p_2\leq p\leq n_2}$ is confined to the set $A_2$ as soon as one the random states $X_{n_1+p_2}\in A_2$ hits the set 
$A_2$ for some $p_2\leq n_2$.   
 
 \begin{equation}\label{MCMC-meth}
 \eta_0\stackrel{M_1^{n_1}}{-\!\!\!\!-\!\!\!\!-\!\!\!\!-\!\!\!\!-\!\!\!\!\longrightarrow} \eta_0 M_1^{n_1}:=\pi_1
 \stackrel{M_2^{n_2}}{-\!\!\!\!-\!\!\!\!-\!\!\!\!-\!\!\!\!-\!\!\!\!\longrightarrow}\pi_1 M_2^{n_2}=\pi_2 \stackrel{M_3^{n_3}}{-\!\!\!\!-\!\!\!\!-\!\!\!\!-\!\!\!\!-\!\!\!\!\longrightarrow}\pi_2 M_3^{n_3}=\pi_3 \stackrel{M_3^{n_3}}{-\!\!\!\!-\!\!\!\!-\!\!\!\!-\!\!\!\!-\!\!\!\!\longrightarrow}\ldots
 \end{equation}

\subsubsection{Feynman-Kac models}
Our next objective is to better understand the evolution of the flow of measures $\eta_p$, from the origin $p=0$ up to the final time $p=n$.
Firstly, it is readily checked that
$$
\PP\left(X\in dx~|~X\in A_{p+1}\right)=\frac{1}{\PP\left(X\in A_{p+1}~|~X\in A_p\right)}~1_{A_{p+1}}(x)~\PP\left(X\in dx~|~X\in A_{p}\right)
$$ 
and 
$$
\PP\left(X\in A_{p+1}~|~X\in A_p\right)=\int 1_{A_{p+1}}(x)~\PP\left(X\in dx~|~X\in A_{p}\right).
$$
  Therefore, in a more synthetic way, if we set $G_p(x)=1_{A_{p+1}}(x)$, then we have that
 $$
  \eta_{p+1}=\Psi_{G_{p}}(\eta_p)
   $$
 with the Boltzmann-Gibbs  $\Psi_{G_{p}}$ transformation  defined by :
 $$
 \eta_p(dx)\longrightarrow \Psi_{G_{p}}(\eta_p)(dx):=\frac{1}{\eta_p(G_p)}~G_p(x)~\eta_p(dx).
 $$
The next formula provides an interpretation of the Boltzmann-Gibbs transformation in terms of
a nonlinear Markov transport equation 
 $$
\Psi_{G_{p}}(\eta_p)(dy)=\left(\eta_p S_{p,\eta}\right)(dy):=\int\eta_p(dx) S_{p,\eta_p}(x,dy)
 $$
 with the Markov transition $ S_{p,\eta_p}$ defined below
 $$
 S_{p,\eta_p}(x,dy)=G_p(x)~\delta_x(dy)+\left(1-G_p(x)\right)~\Psi_{G_{p}}(\eta_p)(dy).
 $$
In summary, we have shown that $(\eta_p)_{0\leq p\leq n}$ satisfies the following evolution equation
  $$
\eta_0\stackrel{S_{0,\eta_0}}{-\!\!\!\!-\!\!\!\!-\!\!\!\!-\!\!\!\!-\!\!\!\!\longrightarrow} \eta_1\stackrel{S_{1,\eta_1}}{-\!\!\!\!-\!\!\!\!-\!\!\!\!-\!\!\!\!-\!\!\!\!\longrightarrow}\eta_2\stackrel{S_{2,\eta_2}}{-\!\!\!\!-\!\!\!\!-\!\!\!\!-\!\!\!\!-\!\!\!\!\longrightarrow}\eta_3\stackrel{S_{3,\eta_3}}{-\!\!\!\!-\!\!\!\!-\!\!\!\!-\!\!\!\!-\!\!\!\!\longrightarrow}\eta_4\ldots
 $$
 In other words, $\eta_p=\mbox{\rm Law}(X^{\star}_p)$ can be interpreted as the laws of the random states of 
 a Markov chain $X^{\star}_p$ with transitions $S_{p,\eta_p}$; that is, we have that
 $$
X^{\star}_0\stackrel{S_{0,\eta_0}}{-\!\!\!\!-\!\!\!\!-\!\!\!\!-\!\!\!\!-\!\!\!\!\longrightarrow} X^{\star}_1\stackrel{S_{1,\eta_1}}{-\!\!\!\!-\!\!\!\!-\!\!\!\!-\!\!\!\!-\!\!\!\!\longrightarrow}X^{\star}_2\stackrel{S_{2,\eta_2}}{-\!\!\!\!-\!\!\!\!-\!\!\!\!-\!\!\!\!-\!\!\!\!\longrightarrow}X^{\star}_3\stackrel{S_{3,\eta_3}}{-\!\!\!\!-\!\!\!\!-\!\!\!\!-\!\!\!\!-\!\!\!\!\longrightarrow}\ldots
 $$
 The Markov chain $X^{\star}_p$ can be interpreted as the optimal sequential acceptance-rejection scheme along the non increasing sequence of subsets $A_{p}$, in the sense that 
 $$
 \left\{
 \begin{array}{rcl}
 X^{\star}_{p}\in A_{p+1}&\Rightarrow&X^{\star}_{p+1}=X^{\star}_{p}\\
  X^{\star}_{p}\in A_{p}-A_{p+1}&\Rightarrow&X^{\star}_{p+1}=X^{\star\star}_{p+1}
 \end{array}
 \right.
 $$
 where $X^{\star\star}_{p+1}$ stand for an independent random sample with distribution $\eta_{p+1}=\Psi_{G_{p}}(\eta_p)$. When the sample 
 $X^{\star}_{p}$ is not in the desired subset $A_{p+1}$, it jumps instantly to a new state $X^{\star\star}_{p+1}$ randomly chosen
 with the desired distribution $\eta_{p+1}=\Psi_{G_{p}}(\eta_p)$. Next we provide a brief discussion on the optimality property of this
 Markov chain model. We recall that
$$
\begin{array}{l}
\|  \eta_{p+1}-  \eta_{p}\|_{\tiny tv}\\
\\
=\sup{\left\{[\eta_{p+1}-  \eta_{p}](f)~:~\mbox{\rm osc}(f)\leq 1 \right\}}\\
\\
=\inf{\left\{
\PP\left(X_p\not=X_{p+1}\right)~:~(X_p,X_{p+1})~~\mbox{s.t.}~~\mbox{\rm Law}(X_p)=\eta_p\quad\mbox{\rm and}\quad \mbox{\rm Law}(X_{p+1})=\eta_{p+1}\right\}}
\end{array}
$$
In the above display $\mbox{\rm osc}(f)=\sup_{x,y}(|f(x)-f(y)|)$ stands for the oscillation of a given function $f$ on $\RR^d$. In this situation, it is instructive to observe that
\begin{equation}\label{coupling-opt}
\|  \eta_{p+1}-  \eta_{p}\|_{\tiny tv}=\PP\left(X^{\star}_{p}\not=X^{\star}_{p+1}\right).
\end{equation}
In other words, the chain $X^{\star}_{p}$ with Markov transitions $S_{p,\eta_p}$ realizes 
the optimal coupling between the sequence of distributions $\eta_p$.
 From the above discussion, we clearly have that
 $$
 \PP\left(X^{\star}_{p+1}\not=X^{\star}_{p}\right)=\eta_p(A_p-A_{p+1})=\eta_p(1-G_p)=1-\eta_p(G_p)
 $$
On the other hand, we have
 \begin{eqnarray*}
  \eta_{p+1}(f)-  \eta_{p}(f)&=&\eta_p\left( S_{p,\eta_p}(f)-f\right)\\
  &=&\eta_p\left([1-G_p]\left[f-\Psi_{G_{p}}(\eta_p)(f)\right]\right)
 \end{eqnarray*}
Choosing $f=1-G_p$, so that $$\Psi_{G_{p}}(\eta_p)(f)=1-\Psi_{G_{p}}(\eta_p)(G_p)=0
$$
and
$$
\eta_p\left([1-G_p]\left[f-\Psi_{G_{p}}(\eta_p)(f)\right]\right)=\eta_p\left([1-G_p]^2\right)=1-\eta_p\left(G_p\right)
$$
This ends the proof of the optimal coupling formulae (\ref{coupling-opt}). Next, we observe that
$$
1-\eta_p\left(G_p\right)=1-\eta_0(A_{p+1})/\eta_0(A_p)
\quad \left(\mbox{\rm with}\quad \eta_0=\mbox{\rm Law}(X)\right)
$$
from which we conclude that
\begin{equation}\label{cond-coupling}
 \eta_0\left(A_{p}\right)\geq  \eta_0\left(A_{p+1}\right)
\geq (1-\epsilon)~\eta_0\left(A_{p}\right)
\Longrightarrow
 \PP\left(X^{\star}_{p+1}=X^{\star}_{p}\right)\geq 1-\epsilon
\end{equation}

As the reader may have certainly noticed, the Markov chain has very poor stability properties, in the sense that 
the distributions $\eta_p$ strongly depends on the initial distribution $\eta_0$. More precisely, $\eta_p$
 coincides with the restriction of $\eta_0$ to the subset $A_p$; more formally, we have that
$$
\eta_p(dx)=\Psi_{G_{p-1}}(\eta_0)=\frac{1}{\eta_0(A_p)}~1_{A_p}(x)~\eta_0(dx)
$$ 

The sequential Monte Carlo methodology is based on combining the MCMC methodology presented (\ref{MCMC-meth}) with the sequential acceptance-rejection technique discussed above. To describe with some precision this method, we let $M_p$ be an MCMC transition with 
invariant measure $\eta_p=\eta_pM_p$.
 In this case, we have the evolution equation
 $$
   \eta_{p+1}=   \eta_{p+1}M_{p+1}=\Psi_{G_{p}}(\eta_p)M_{p+1}:=\Phi_{p+1}(\eta_p)
 $$
 Notice that $\Phi_{p+1}$ maps the set of probability measures $\eta$ s.t. $\eta(G_p)>0$ into the set of probability measures, and it is the composition of an updating transformation $\Psi_{G_{p}}$ and a Markov transport equation w.r.t. $M_{p+1}$; that is, we have that
 $$
\eta_p \stackrel{\Psi_{G_p}}{-\!\!\!\!-\!\!\!\!-\!\!\!\!-\!\!\!\!-\!\!\!\!\longrightarrow}\widehat{\eta}_p:=\Psi_{G_{p}}(\eta_p)\stackrel{M_{p+1}}{-\!\!\!\!-\!\!\!\!-\!\!\!\!-\!\!\!\!-\!\!\!\!\longrightarrow}\widehat{\eta}_pM_{p+1}=\Phi_{p+1}(\eta_p)
 $$
 The solution of this equation is given by the Feynman-Kac measures defined for any measurable
 function
 $f$ on $\RR^d$ by the following formulae
\begin{equation}\label{FK-model}
\eta_p(f)={\gamma_p(f)}/{\gamma_p(1)}\quad\mbox{\rm with}\quad
\gamma_p(f)=\EE\left(f(X_p)~\prod_{0\leq q<p}G_q(X_q)\right).
\end{equation}
To prove this claim, we use the Markov property to check that
\begin{eqnarray*}
 \gamma_{p+1}(f) &=&\EE\left(M_{p+1}(f)(X_{p})~G_{p}(X_{p})\prod_{0\leq q<p}G_q(X_q)\right)=\gamma_p(G_pM_{p+1}(f)).
 \end{eqnarray*}
 This clearly implies that
 $$
 \eta_{p+1}(f) =\frac{\gamma_p(G_pM_{p+1}(f))/\gamma_p(1)}{ \gamma_p(G_p)/\gamma_p(1)}=\frac{\eta_p(G_pM_{p+1}(f))}{\eta_p(G_p)}=\Psi_{G_{p}}(\eta_p)M_{p+1}(f).
 $$

We already mention that the unnormalized measures $\gamma_n$ can be expressed in terms of the flow of measures $(\eta_p)_{0\leq p\leq n}$ with the following multiplicative formulae
\begin{equation}\label{UFK-model}
 \gamma_p(f)=\eta_p(f)\times \prod_{0\leq q<p}\eta_q(G_q)
\end{equation}
 This result is a direct consequence of the following observation
 $$
 \gamma_{p}(1)=\EE\left(G_{p-1}(X_{p-1})\prod_{0\leq q<p-1}G_q(X_q)\right)=\gamma_{p-1}(G_{p-1})=\eta_{p-1}(G_{p-1})~\gamma_{p-1}(1).
 $$
 
It is readily checked that the measures $\eta_n$ are the $n$-th time marginals of the Feynman-Kac measures on path space defined by
the following formulae 
\begin{equation}\label{Q-def}
 d\mathbb{Q}_{n}:=\frac{1}{\mathcal{Z}_{n}}~\left\{ \prod_{0\leq
p<n}G_{p}(X_{p})\right\} ~d\mathbb{P}_{n}
\end{equation}
with some normalizing constants $\mathcal{Z}_{n}=\gamma_n(1)$ and the reference measures
$$
\mathbb{P}_{n}=\mbox{\rm Law}(X_0,\ldots,X_n).
$$ 
This class of path space measures goes beyond the MCMC model discussed above. These measures represent the distribution of the trajectories of a reference Markov process, weighted by a collection of potential functions. These functional models are natural mathematical extensions of the traditional change of probability measures, commonly used in importance sampling. 

From a pure probabilistic viewpoint, these measures can be interpreted as the conditional distribution of a given Markov chain w.r.t. to a sequence of events. For instance, if we take $G_n=1_{A_n}$ indicator potential functions of some measurable subsets $A_n\in E_n$, then we readily check that
$$
\QQ_n=\mbox{\rm Law}((X_0,\ldots,X_n)~|~\forall 0\leq p<n~~X_p\in A_p)
~~\mbox{\rm and}~~
 \Za_n=\PP\left(\forall 0\leq p<n~~X_p\in A_p\right)
 $$
In filtering settings, if we take $G_n(x_n)=p(y_n|x_n)$ the likelihood function associated with the observation $Y_n=y_n$ of the random signal state
$X_n$, then we have
 $$
 \QQ_n=\mbox{\rm Law}((X_0,\ldots,X_n)~|~\forall 0\leq p<n~~Y_p=y_p)\quad\mbox{\rm and}\quad
 \Za_n=p(y_0,\ldots,y_{n-1})
 $$
 For a more thorough discussion on the application domains of these Feynman-Kac models, we refer the reader to the books
~\cite{cappe,dm04,dm12,Doucet:2001}.
 \subsection{Nonlinear distribution flows}
 The central idea behind Feynman-Kac particle samplers is to observe that {\em any} evolution equation of probability measures
$$
\eta_n=\Phi_{n}\left(\eta_{n-1}\right)
$$
on some measurable state spaces $E_n$
can be interpreted as the law 
$$
\eta_n=\mbox{\rm Law}\left(\overline{X}_n\right)
$$
of a Markov chain $\overline{X}_n$ with initial distribution $\eta_0$ and Markov transitions
$$
\PP\left(\overline{X}_n\in dx_n~|~\overline{X}_{n-1}=x_{n-1}\right)=K_{n, \eta_{n-1}}(x_{n-1},dx_n).
$$ 
The Markov transitions $K_{n,\eta_{n-1}}$ are chosen so that
 $$
\forall n\geq 1\qquad \eta_{n-1}K_{n, \eta_{n-1}}=\Phi_n(\eta_{n-1}).
 $$
 The Markov chain $\overline{X}_n$ incorporate free evolution moves according to $M_n$, with sequential updates of the measures $\eta_n$, so that the law of the random states $\overline{X}_n$ coincide with the desired distributions $\eta_n$, at every time step.
In this interpretation, this chain
can be interpreted as a perfect sequential sampler of the  sequence of measures $\eta_n$.

The choice of the transitions $K_{n+1,\eta_n}$ is not unique. For instance, for the Feynman-Kac models on $E_n=\RR^d$ discussed above, if we take
$$
K_{n+1,\eta_n}(x,dy):=\left[S_{n,\eta_n} M_{n+1}\right](x,dy)\quad\mbox{\rm or}\quad
K_{n+1,\eta_n}(x,dy):=\Phi_{n+1}\left(\eta_n\right)(dy)
$$
we readily check that
$$
\eta_n K_{n+1,\eta_n}=\Phi_{n+1}\left(\eta_n\right)= \Psi_{G_{n}}(\eta_n)M_{n+1}= \eta_n S_{n,\eta_n} M_{n+1}.
$$
We also mention that the law of the random trajectories $(\overline{X}_0,\ldots,\overline{X}_n)$ are given by the so-called McKean measures
$$
\overline{\PP}_n(d(x_0,\ldots,x_n))=\eta_0(dx_0)~K_{1,\eta_0}(x_0,dx_1)\ldots K_{n,\eta_{n-1}}(x_{n-1},dx_{n}).
$$

We further assume that the Markov transitions $M_n(x_{n-1},dx_n)$ are absolutely continuous with respect to some reference measure $\nu_n$ and we set
\begin{eqnarray*}
Q_n(x_{n-1},dx_n):=G_{n-1}(x_{n-1})M_n(x_{n-1},dx_n)=
H_n(x_{n-1},x_n)~\nu_n(dx_n).
\end{eqnarray*}
In this situation, we have the following time reversal formulae
\begin{equation}\label{backward}
 \QQ_n(d(x_0,\ldots,x_n))=\eta_n(dx_n)~\MM_{n,\eta_{n-1}}(x_n,dx_{n-1})\ldots \MM_{1,\eta_{0}}(x_1,dx_{0})
\end{equation}
with the Markov transitions 
$$
\MM_{n,\eta_{n-1}}(x_n,dx_{n-1}):=\frac{\eta_{n-1}(dx_{n-1})~H_{n}(x_{n-1},x_{n})}{\eta_{n-1}\left(H_{n}(\point,x_{n})\right)}.
$$
We prove this backward formula using the fact that
$$
\eta_n(dx_n)=\Psi_{G_{n-1}}(\eta_{n-1})M_{n}(dx_n)=\frac{\eta_{n-1}\left(H_{n}(\point,x_{n})\right)}{\eta_{n-1}(G_{n-1})}~~\nu_n(dx_n)
$$
from which we find that
$$
\eta_n(dx_n)~\MM_{n,\eta_{n-1}}(x_n,dx_{n-1})=\frac{1}{\eta_{n-1}(G_{n-1})}~\eta_{n-1}(dx_{n-1})~ Q_n(x_{n-1},dx_n).
$$
Iterating this process, we prove (\ref{backward}).
 \subsubsection{Interacting particle methods}
 This section is concerned with particle approximations of the Feynman-Kac model (\ref{FK-model}) and (\ref{Q-def}).
 We also present a series of exponential concentration inequalities that allows to estimate the deviation of 
the particle estimates around their limiting values. 

In the further development of this section $f_n$ stands for some function s.t. $\|f_n\|\leq 1$, and
 $(c_1,c_2)$ represent two constants related to the bias and the variance of the particle approximation scheme, and $c$ stands for some universal constant. The values of these constants may vary from line to line but they don't depend on the time horizon. 
Last, but not least, we assume that the Feynman-Kac model satisfies some strong stability properties. 
 For a more detailed description of the stability properties, and the description of the quantities $(c,c_1,c_2)$ in terms
 of the Feynman-Kac model (\ref{FK-model}), we refer the reader to the books~\cite{dm04,dm12}.

We approximate the transitions 
$$
\overline{X}_{n}\leadsto \overline{X}_{n+1}\sim K_{n+1,\eta_{n}}(\overline{X}_{n},dx_{n+1})
$$
 by running a Markov chain $\xi_n=(\xi^1_n,\ldots,\xi^N_n)\in E_n^N$ that approximate the distribution $\eta_n$ when $N\uparrow\infty$
$$\frac{1}{N}\sum_{1\leq i\leq N}\delta_{\xi^i_n}:=\eta^{N}_n~\longrightarrow_{N\uparrow\infty}~ \eta_n.
$$
A natural choice of particle transitions is to take at every time step sequence of conditionally independent particles
$$
\xi^i_{n}\leadsto \xi^i_{n+1}\sim K_{n+1,\eta_{n}^N}(\xi^i_{n},dx_{n+1}).
$$
For the Feynman-Kac models discussed above, we can chose the transitions $K_{n+1,\eta_n}=S_{n,\eta_n} M_{n+1}$. In this context,
the evolution of the particle algorithm is decomposed into two steps.
 $$
 \left.
 \begin{array}{cl}
 \xi^1_{n}&\\
\vdots&\\
 \xi^i_{n}&\\
\vdots&\\ 
  \xi^N_{n}&
 \end{array}
 \right]\stackrel{S_{G_n,\eta^N_n}}{-\!\!\!\!-\!\!\!\!-\!\!\!\!-\!\!\!\!-\!\!\!\!-\!\!\!\!-\!\!\!\!-\!\!\!\!-\!\!\!\!\longrightarrow }
  \left[
 \begin{array}{cl}
\widehat{\xi}^1_{n}&\stackrel{M_{n+1}}{-\!\!\!\!-\!\!\!\!-\!\!\!\!-\!\!\!\!-\!\!\!\!-\!\!\!\!-\!\!\!\!-\!\!\!\!-\!\!\!\!\longrightarrow}\\
\vdots&\\
\widehat{\xi}^i_{n}&-\!\!\!\!-\!\!\!\!-\!\!\!\!-\!\!\!\!-\!\!\!\!-\!\!\!\!-\!\!\!\!-\!\!\!\!-\!\!\!\!\longrightarrow\\
\vdots&\\ 
 \widehat{\xi}^N_{n}&-\!\!\!\!-\!\!\!\!-\!\!\!\!-\!\!\!\!-\!\!\!\!-\!\!\!\!-\!\!\!\!-\!\!\!\!-\!\!\!\!\longrightarrow
 \end{array}
 \right.
 \left.
 \begin{array}{cl}
 \xi^1_{n+1}&\\
\vdots&\\
 \xi^i_{n+1}&\\
\vdots&\\ 
  \xi^N_{n+1}&
 \end{array}
 \right]
 $$
During the first step, every particle $\xi^i_{n}$ evolves to a new particle $\widehat{\xi}^i_{n}$ randomly chosen with the distribution
 $$
 S_{\eta^N_{n}}( \xi^i_{n},dx)
 :=G_n( \xi^i_{n})~\delta_{ \xi^i_{n}}(dx)+
 \left(1-G_n( \xi^i_{n})\right)~ \Psi_{G_n}(\eta^N_n)(dx)
  $$
 with the updated measures
 $$
 \Psi_{G_n}(\eta^N_n)=\sum_{j=1}^N\frac{G_n( \xi^j_{n})}{\sum_{k=1}^NG_n( \xi^k_{n})}
 \delta_{ \xi^j_{n}}
\longrightarrow_{N\uparrow\infty} \Psi_{G_n}(\eta_n)=\eta_{n+1}.
 $$
This transition can be interpreted as an acceptance-rejection scheme with a recycling mechanism. In the second step, the selected particles
$\widehat{\xi}^i_{n}$ evolve randomly according to the Markov transitions $M_{n+1}$. In other words, for any $1\leq i\leq N$, we sample
a  random state $ \xi^i_{n+1}$ with distribution $M_{n+1}\left(\widehat{\xi}^i_{n},dx\right)$.

Using the concentration analysis of mean field particle models developed in~\cite{dmr-11}, the following exponential estimate was proved in~\cite{dm12}.For any $x\geq 0$, $n\geq 0$, and any population size $N\geq 1$, the probability of the event
$$
\left[\eta^N_{n}-\eta_n\right](f)\leq \frac{c_1}{N}~\left(1+x+\sqrt{x}\right)+\frac{c_2}{\sqrt{N}}~\sqrt{x}
$$
is greater than $1-e^{-x}$. In addition,
for any $x=(x_i)_{1\leq i\leq d}$ and any $(-\infty,x]=\prod_{i=1}^d(-\infty,x_i]$ cells in $E_n=\RR^d$, we let
$$
F_n(x)=\eta_n\left(1_{(-\infty,x]}\right)\quad\mbox{\rm and}\quad
F^N_n(x)=\eta^N_n\left(1_{(-\infty,x]}\right).
$$
For any $y\geq 0$, $n\geq 0$, and any population size $N\geq 1$, the probability of the following event
 $$
\sqrt{N}~\left\|F^N_n-F_n\right\|\leq c~\sqrt{d~(y+1)}
$$
is greater than $1-e^{-y}$.

If we interpret the mutation-selection particle algorithm
as a birth and death branching process, then we can trace back in time
the whole ancestral line $\zeta^i_n=(\xi_{p,n}^{i})_{0\leq p\leq n}$ of the individual $\xi^i_n$ at the $n$-th
generation.
$$
\xi_{0,n}^{i}\longleftarrow\xi_{1,n}^{i}\longleftarrow\ldots\longleftarrow
\xi_{n-1,n}^{i}\longleftarrow\xi_{n,n}^{i}=\xi_{n}^{i}
$$ 
The random state $\xi^i_{p,n}$ represents the ancestor of the
individual $\xi^i_n$ at the level $p$, with $0\leq p\leq n$, and
$1\leq i\leq N$. It is more or less well known that 
$\zeta_n$ coincides with the particle approximation of the Feynman-Kac model
defined as in (\ref{FK-model}) by replacing $X_n$ by the historical process $(X_p)_{0\leq p\leq n}$.
This interpretation provides an alternative particle approximation scheme of the measures (\ref{Q-def}), that is we have that
$$
\eta_n^N=\frac{1}{N}\sum_{1\leq i\leq N}\delta_{\left(\xi^i_{0,n},\xi^i_{1,n},\ldots,\xi_{n,n}^i\right)}
~\longrightarrow_{N\uparrow\infty}~ \QQ_n
$$
More precisely, we proved in~\cite{dm12} the following exponential concentration estimate.
For any test function ${\bf f_n}$ on path space s.t. $\|{\bf f_n}\|\leq 1$,
for any $y\geq 0$, $n\geq 0$,  and any $N\geq 1$,  the probability of the event
$$
\left[\eta^N_{n}-\QQ_n\right](f)\leq c_1~\frac{n+1}{N}~\left(1+x+\sqrt{x}\right)+c_2~\sqrt{\frac{(n+1)}{N}}~\sqrt{x}
$$
is greater than $1-e^{-x}$.

Further details on these genealogical tree models can be found in~\cite{dm04,dm12,dmm-01}.
Mimicking formulae (\ref{UFK-model}) and  (\ref{backward}), we define {\em an unbiased} particle estimate $\gamma^N_n$ of the unnormalized measures $\gamma_n$ and a particle backward measures $\QQ^N_n$
by setting
$$
  \gamma_n^N(f)=\eta_n^N(f)\times \prod_{0\leq q<n}\eta_q^N(G_q)
  $$
and
$$
 \QQ^N_n(d(x_0,\ldots,x_n))=\eta_n^N(dx_n)~\MM_{n,\eta^N_{n-1}}(x_n,dx_{n-1})\ldots \MM_{1,\eta^N_{0}}(x_1,dx_{0})
$$
We end this section with a couple of   exponential concentration estimates proved in~\cite{dm12} .
 For any $x\geq 0$, $n\geq 0$, $N\geq 1$, and any $\epsilon\in \{+1,-1\}$, the probability of the event
$$
\frac{\epsilon}{n}\log{\frac{\gamma^N_n(1)}{\gamma_n(1)}}
\leq  \frac{c_1}{N}~\left(1+x+\sqrt{x}\right)+\frac{c_2}{\sqrt{N}}~\sqrt{x}
$$
is greater than $1-e^{-x}$. In addition, for any
normalized additive functional ${\bf f_n}(x_0,\ldots,x_n)=\frac{1}{n+1}\sum_{0\leq p\leq n}f_p(x_p)$ with $\|f_p\|\leq 1$, 
for $x\geq 0$, $n\geq 0$, and any population size $N\geq 1$, the probability of the event
$$
\left[\QQ ^N_n-\QQ_n\right](\overline{\bf f}_n)
\leq \displaystyle c_1~\frac{1}{N}~(1+(x+\sqrt{x}))+
c_2~\sqrt{\frac{x}{N(n+1)}}
$$
is greater than $1-e^{-x}$.

\section{Illustration of Interacting Particle Solutions for Risk and Insurance Capital Estimation}
\label{SectionExample}
In this section we detail a special sub-set of algorithms from within the stochastic particle integration methods that was specifically developed to solve problems for risk and insurance in \cite{peters2007simulation} and discussed in comparison to specific FFT methods in \cite{luo2009computing}. The class of recursive solutions developed is very general and applicable to a wide range of insurance and risk settings. We provide a novel result in this illustration which extends the framework originally presented in \cite{peters2007simulation} through consideration of a higher-order recursion based on the widely utilized Panjer recursion (avoiding the need to perform discretization of the severity distribution).

\textit{It is important to understand that whilst this method is focusing on asymptotic results for tail functionals of a compound processes model under the LDA framework and can be re-engineered into a class of rare-event type algorithm as discussed in Section \ref{Section:SMCMethods}, the presentation of this example that we adopt will be more general and therefore more applicable to a wide class of insurance and risk problems that include but is not limited to those used to motivate the use of stochastic particle solutions in the introduction.}

Finally, we point out that this method to be discussed is one of many methods that could be developed for such an insurance problem, though it has proven to be highly efficient even in the simplest form of the sampler as discussed below. In addition we note that as discussed, whilst the asymptotic approximation methods that were presented were developed to tackle the serious statistical and computational challenges posed by accurate estimation of tail quantiles and expectations for heavy-tailed LDA models. Noting that their elegance lies in the fact they allow one to bypass the serious computational challenges for estimation of risk measures for such heavy-tailed annual loss distributions under traditional integration methods, Fourier methods, recursions (Panjer) or basic Monte Carlo approaches, they do have associated issues, see discussions in \cite{hess2011can} and the arguments presented in the introduction of this article. As discussed the properties of such SLA estimates is still an active subject of study with regard to approximation error, unbiased quantile function estimation, asymptotic rates of convergence, sensitivity to parameter estimation and model misspecification. Hence, one often requires calculation of VaR, ES and Spectral Risk Measures which do not utilize such asymptotic properties but instead are based on stochastic integration methods. Here we consider such an example.

\subsection{Recursions for Loss Distributions: Panjer and Beyond}
The framework proposed in \cite{peters2007simulation} and \cite{doucet2010solving} for developing a recursive numerical solution to estimation of such risk measures through estimation of the density of the compound process. In particular we briefly summarize an approach to transform the standard actuarial solution known as the Panjer recursion \cite{panjer1981recursive} to a sequence of expectations. Recursions for the evaluation of single risk process distributions are ubiquitous in risk and insurance modeling, see the book length review in discrete loss distributional settings in \cite{sundt2009recursions}. In this paper we consider an advanced development that avoids the need to discretize the severity distribution to perform inference and instead develops a stochastic particle integration based solution, hence providing a fitting example for such methods in risk and insurance.

Recursions for evaluating $\overline{F}_{Z}(x)$ and risk measures for Subexponential LDA models are traditionally in risk modeling based around in the most fundamental class of models, the Panjer class
Panjer Recursion Class of frequency distribution relationships defined by,
\begin{equation}
p_{n}=\left( a+\frac{b}{n}\right) p_{n-1}
\end{equation}
with members Poisson $(a=0,b=l,p_{0}=e^{-\lambda })$, Binomial $(a=\frac{-q}{(1-q)},b=\frac{(m+1)q}{(1-q)},p_{0}=(1-q)m)$ and Negative Binomial $(a=\frac{b}{1+b},b=\frac{(r-1)b}{1+b},p_{0}=(1+b)-r) $.
In addition, one can can derive the generalized (higher order Panjer) recursion for an extended class of frequency distributions given by the generalized Poisson distribution (GPD) via
\begin{equation*}
\mathbb{P}\mathrm{r}\left(N=n\right) = p_n(\lambda,\theta) = 
\begin{cases}
\lambda(\lambda + n\theta)^{n-1}\frac{}{}, \; &\forall n = 0,1,2,\ldots \\
0, \; & \mathrm{if} \; n > m, \; \mathrm{when} \; \theta < 0.
\end{cases}
\end{equation*}
with $\lambda > 0$ and $\max(-1,\lambda/m) \leq \theta < 1$ and $m \geq 4$ is the largest positive integer s.t. $\lambda + \theta m > 0$ when $\theta$ is negative, where the GPD is Poisson for $\theta = 0$; over-dispersed $\theta > 0$ and under-dispersed $\theta < 0$.

Considering these classes one can derive closed form recursions for the annual loss LDA compound process distribution according to the Panjer recursion \cite{panjer1981recursive},
\begin{equation}
f_{Y}\left( x \right) = p_{1}f_{X}\left( x\right) + \int_{0}^{x}\left(
a+\frac{by}{x}\right) f_{X}\left( y\right) f_{Y}\left( x-y\right) dy
\end{equation}
or the generalized higher order Panjer recursion \cite{goovaerts1991evaluating},
\begin{equation}
f_{Y}\left( x \right) = p_{1}(\lambda,\theta)f_{X}\left( x\right) + \frac{\lambda}{\lambda + \theta} \int_{0}^{x}\left(
\theta+ \lambda \frac{y}{x}\right) f_{X}\left( y\right) f_{Y}\left(x-y\right) dy
\end{equation}
To understand where these recursions are obtained from, consider the convolution identity for an i.i.d. partial sum $S_{n+1} = X_1 + \ldots + X_{n+1}$ with density
\begin{equation}
f^{*(n+1)}(x) = \int_{0}^{x} f(\tau)f^{*n}(x-\tau)d\tau, \;\; \forall n = 1,2,3,\ldots.
\end{equation}
Substitute the conditional of $X_1$ when $S_{n+1} = x$,
\begin{equation}
f_{X_1}\left(\tau | X_1 + \cdots + X_{n+1} = x\right) = \frac{f(\tau)f^{*n}(x-\tau)}{f^{*(n+1)}(x)}.
\end{equation}
into the average given $S_{n+1} = x$ to get
\begin{equation} 
\mathbb{E}\left[X_1 | X_1 + \cdots + X_{n+1} = x\right] = \int_{0}^{x} \tau \frac{f_{X_1}(\tau)f^{*n}_{X_1}(x-\tau)}{f^{*(n+1)}_{X_1}(x)} d\tau.
\end{equation}
Then observe that with i.i.d. losses one also gets
\begin{equation}
\begin{split}
&\mathbb{E}\left[X_1 | X_1 + \cdots + X_{n+1} = x\right] = \frac{1}{n+1} \sum_{i=1}^{n+1}\mathbb{E}\left[X_i | X_1 + \cdots + X_{n+1} = x\right] \\
&= \frac{1}{n+1}\mathbb{E}\left[X_1 + \cdots + X_{n+1} | X_1 + \cdots + X_{n+1} = x\right] = \frac{x}{n+1}
\end{split}
\end{equation}
Equating these conditional mean expressions and rearranging gives 
\begin{equation}
\frac{1}{n+1}f^{*(n+1)}_{X_1}(x) = \frac{1}{x} \int_{0}^{x} \tau f_{X_1}(\tau)f^{*n}_{X_1}(x-\tau) d\tau.
\end{equation}
Now utilize the Panjer class of frequency distributions satisfying for some $a$ and $b$,
\begin{equation}
\mathbb{P}\mathrm{r}\left(N=n\right)= p_n = \left(a + \frac{b}{n}\right)p_{n-1},
\end{equation}
Upon substitution and some elementary algebra one obtains the Panjer recursion.

\subsection{Stochastic Particle Integration Methods as Solutions to Panjer Recursions}
Solving these recursions is then typically performed to obtain estimates of $\overline{F_{Z_N}}(x)$. In the case of sub-exponential severity models this can be costly if the standard actuarial approach is adopted which involves discretization of the severity distribution. This renders the integral equation in a form for certain classes of frequency and severity models that admit the De Pril transform solution or approximations of such a transform solution, see discussion in \cite{sundt2009recursions}. Such approaches are then prone to a trade-off between discretization errors and computational efficiency, though the errors are deterministic. The second approach one may consider is to avoid the discretization error in favor of a Monte Carlo solution. If this can be done efficiently with a clear understanding of the associated Monte Carlo errors, this may be considered as a preferred alternative. Such an approach was adopted under a path space based Importance Sampling solution in \cite{peters2007simulation}.

It was noted in \cite{peters2007simulation} that since the Panjer recursions could be re-expressed as linear Volterra integral equations of the second kind via the mapping 
\begin{equation}
\begin{split}
&x_1 = x - y, \; g(x) = p_1 f_X(x), \; f\left(x_1\right) = f_{Z}(x_1), \, \mathrm{and}\\
&k\left(x,x_1\right) = \left(a + b\frac{x - x_1}{x}\right)f_X\left(x-x_1\right).
\end{split}
\end{equation}
where the kernel $k:E \times E \mapsto \mathbb{R}$ and function $g:E \mapsto \mathbb{R}$ are known and $f:E \mapsto \mathbb{R}$ is unknown. Furthermore, if one defines $k^0(x,y) \triangleq 1$, $k^1(x,y) \triangleq k(x,y)$ and 
$$
k^n(x,y) \triangleq \int k(x,y)k^{n-1}(z,y)dz
$$
and these kernels satisfy that
$$
\sum_{n=0}^{\infty} \int_{E}\left|k^n\left(x_0,x_n\right)g\left(x_n\right)\right|dx_n < \infty
$$
then one can identify the resolvent kernel and Neumann series through iterative expansion of the recursion to obtain for a sequence of domains $E_{1:n}$
\begin{equation*}
f\left(x_0\right) = g(x_0) + \sum_{n=0}^{\infty} \int_{0}^{x_0} \ldots \int_{0}^{x_{n-1}} g\left(x_n\right) \prod_{l=1}^{n} k\left(x_{l-1},x_{l}\right) dx_{1:n},
\end{equation*}
see further details in \cite{peters2007simulation}. Under this formulation it was shown in \cite{peters2007simulation} how to address two problems: \textsl{ estimation of the annual loss density over a set $A$ and estimation of the annual loss density point wise}. These are both directly relevant to obtaining estimates of the risk measures specified. 

To proceed one converts the Neumann series into a sequence of expectations with respect to an importance sampling distribution given by associating the following elements
\begin{equation*}
\begin{split}
f_0\left(x_0\right) &= g\left(x_0\right), \; \mathrm{and} \; f_n\left(x_{0:n}\right) = g\left(x_n\right) \prod_{l=1}^n k\left(x_{l-1},x_l\right)\\
\therefore\, f\left(x_0\right) &= f_0\left(x_0\right) + \sum_{n=1}^{\infty}\int_{0}^{x_0}\ldots \int_{0}^{x_{n-1}} f_n\left(x_{0:n}\right) dx_{0:n}.
\end{split}
\end{equation*}
in order to frame this problem as an expectation with respect to a sequence of distributions $\left\{\pi\left(n,x_{1:n}\right)\right\}_{n \geq 0}$:
\begin{equation*}
\begin{split}
f(x) &= \frac{f_0(x)}{\pi(0)}\pi(0) + \sum_{n=1}^{\infty}\int_{A_{1:n}(x)}\frac{f_n\left(x,x_{1:n}\right)}{\pi\left(n,x_{1:n}\right)}\pi\left(n,x_{1:n}\right)dx_{1:n} \\
&=\mathbb{E}_{\pi\left(n,x_{1:n}\right)}\left[\frac{f_n\left(x,x_{1:n}\right)}{\pi\left(n,x_{1:n}\right)}\right],
\end{split}
\end{equation*}
with the sets $A_{1:n}\left(x_0\right) = \left\{\left(x_1,\ldots,x_n\right):\, x_0 > x_1 > \cdots > x_n \right\}$ playing an analogous role to the sequence of level sets described in the methodology section of the paper.

Hence we note there are clearly now two path-space based particle solutions available, those that consider estimating $f(x)$ point-wise which involves an importance sampling solution on the path-space defined by 
$$ \cup_{n=0}^{\infty} \left\{n\right\}\times A_{1:n}(x).$$ 
Note, efficient variance reduction (per sample basis and per unit of computation basis) involves instead estimating $\left(f(x) - f_0\left(x_0\right) \right)$ by importance sampling on smaller space 
$$\cup_{n=1}^{\infty} \left\{n\right\} \times A_{1:n}(x).$$
The other alternative involves characterizing $f(x)$ over some interval by obtaining samples from its restriction to that interval $\left[x_a,x_b\right]$, via importance sampling on a slightly larger space 
$$\bigcup_{n=0}^{\infty} \left\{n\right\}\times A_{1:n}\left(\left[x_a,x_b\right]\right).$$

One can now consider constructing importance sampling based solutions to this sequence of expectations as detailed in \cite{peters2007simulation} [Algorithm 1, p.9] and [Algorithm 2, p.12] and \cite{doucet2010solving} [Algorithm 2.1.1] and for which we provide the relevant pseudo code in Algorithm 2. This is summarized according to the path-space based Sequential Importance Sampling (SIS) based approximation to the annual loss distribution, for a Markov chain with initial distribution/density $\mu(x) > 0$ on $E$ and transition kernel $M(x,y)>0$ if $k(x,y) \neq 0$ and $M$ has absorbing state $d \notin E$ such that $M(x,d) = P_d$ for any $x \in E$, by the following steps in Algorithm 2.\\
\\
\noindent \textbf{Algorithm 2: Stochastic Particle Methods on Path-Space for Panjer Recursions}
\begin{enumerate}
\item{Generate $N$ independent Markov chain paths $\left\{X^{(i)}_{0:n^{(i)}+1}\right\}_{i=1}^N$ until absorption $X^{(i)}_{n^{(i)}+1}=d$.}
\item{Evaluate the importance weights for each particle on the path space by,
\begin{equation}
W\left(X^{(i)}_{0:n^{(i)}}\right) = \begin{cases}
\frac{1}{\mu\left(X^{(i)}_{0}\right)}\left(\prod_{n=1}^{n^{(i)}}\frac{k\left(X^{(i)}_{n-1},X^{(i)}_{n}\right)}{M\left(X^{(i)}_{n-1},X^{(i)}_{n}\right)} \right) \frac{g\left(X^{(i)}_{n^{(i)}}\right)}{P_d}, & \mathrm{if } n^{(i)} \geq 1,\\
\frac{g\left(X^{(i)}_{0}\right)}{\mu\left(X^{(i)}_{0}\right)P_d}, & \mathrm{if } n^{(i)} = 0.
\end{cases}
\end{equation}
}
\end{enumerate}

\subsection{Stochastic Particle Solutions to Risk Measure Estimation}
Then if $\mu\left(X^{(i)}_{0}\right) = \delta\left(X^{(i)}_{0}\right), \, \forall i \in \left\{1,\ldots,N \right\}$, then the empirical measure at \textit{a point $x_0$} is given by: 
$$
\widehat{f}_Z\left(x_0\right) = \frac{1}{N}\sum_{i=1}^N W\left(x_0,X^{(i)}_{1:n^{(i)}}\right)
$$
or over an interval,
$$
\widehat{f}_Z\left(x_0\right) = \frac{1}{N}\sum_{i=1}^N W_1\left(X^{(i)}_{0:n^{(i)}}\right) \delta\left(x_0 - X_0^{(i)}\right)
$$
forms an unbiased Monte Carlo approximation of the expectation of $f_Z(z)$ for any set $A$ given by $\mathbb{E}\left[\int_{A}\widehat{f}(x_0) dx_0\right] = \int_{A}f(x_0) dx_0$. Furthermore, detailed discussions on the optimal choice with respect to minimizing the variance of the importance weights is developed in \cite{peters2007simulation} and \cite{doucet2010solving}. 

Having obtained this particle based approximation, this weighted dirac measure can then be utilized to estimate any of the required risk measures such as VaR, ES and SRM for any desired level $\alpha$. This can be performed in two ways, depending on whether the particle solution is obtained for the evaluation of the recursions point wise over a fixed grid or alternatively over an interval, which could be increasing in size. In the case of an interval, or contracting set, one considers perhaps a set of interest to be $A = [0,x_{\max}]$ such that $x_{max} >> F^{\leftarrow}\left(1-\frac{1-\alpha}{\mathbb{E}[N]}\right)$, and then utilize this to construct an unbiased particle approximation of the distribution of the annual loss up to any level $\alpha \in (0,1)$. Clearly, this could be obtained from growing a set $A_1 = [0,x_1] \subset A_2 \subset \cdots \subset A= [0,x_{\max}]$ recursively, as discussed in previous sections where such as set would result in the tail $\overline{F}$ being restricted to a smaller and smaller set of the complement rare event. 

If the partition is made point wise over a linear or a non-linear spacing $[0,z] = \bigcup_{m=1}^M \left[(m-1)\triangle,m\triangle\right)$ and the distribution evaluated point-wise, this leads to an estimation of
\begin{equation}
\begin{split}
\widehat{F}_Z(z) &= \sum_{m=0}^M \triangle f\left(m\triangle\right) 
\approx \frac{1}{N}\sum_{m=0}^M \sum_{i=1}^N \triangle W\left(m\triangle,X^{(i,m)}_{1:n^{(i,m)}}\right). 
\end{split}
\end{equation}
Alternatively, if the estimation is performed over an interval, this construction of the empirical measure over interval $A(x_{\max}) = \left[0,x_{\max}\right]$ such that for any $z < x_{\max}$ results in,
\begin{equation}
\widehat{F}_Z\left(z\right) = \frac{1}{N}\sum_{i=1}^N W\left(X^{(i)}_{0:n^{(i)}}\right) \mathbb{I}\left(X_{0:n^{(i)}}^{(i)} \in [0,z] \right) \rightarrow_{N \uparrow \infty} \int_{0}^{z} f_{Z}(z) dz. 
\end{equation}
\textbf{Practical advice:} consider a range for the support $[0,x_{\max}]$ s.t. $x_{max} >> F^{\leftarrow}\left(1-\frac{1-\alpha}{\mathbb{E}[N]}\right)$.

From these unbiased particle approximations of the annual loss density and distribution, the evaluation of the risk measures for VaR, ES and SRM follows trivially. To begin with the VaR or indeed any quantile, it would be useful to have a reconstruction of the quantile function of the annual loss LDA model distribution function. The quantile function is defined by
\begin{equation}
Q(p) = F_Z^{\leftarrow}(p) = \inf \left\{x \in \mathbb{R}: \, p \leq F_Z(x) \right\}.
\end{equation}
Estimation of the Quantile function can now be obtained by approximation utilizing the empirical measure either based on a random set of particle locations or a discrete deterministic grid as follows:
\begin{itemize}
\item[]{\textbf{Deterministic Grid Solution:} Given partition $[0,x_{\max}] = \bigcup_{m=1}^M \left[(m-1)\triangle,m\triangle\right)$ for some step $\triangle$ s.t.
\begin{equation}
\widehat{Q}(p) = \inf \left\{x \in \left\{0,\triangle,\ldots,M\triangle\right\}: \, p \leq
\frac{1}{N}\sum_{m=0}^M \sum_{i=1}^N \triangle W\left(x,X^{(i,m)}_{1:n^{(i,m)}}\right). 
\right\}.
\end{equation}
Guide - set $x_{max} >> F^{\leftarrow}\left(1-\frac{1-\alpha}{\mathbb{E}[N]}\right)$.
}
\item[]{\textbf{Interval Solution:} Construct the empirical measure over $A(\infty) = \left[0,\infty\right)$ s.t.
\begin{equation}
\widehat{Q}(p) = \inf \left\{x \in \left\{X_{(0)}^{(i)}\right\}_{i=1:N}: \, p \leq \frac{1}{N}\sum_{i=1}^N W\left(X^{(i)}_{(0):n^{(i)}}\right) \mathbb{I}\left(X_{(0):n^{(i)}}^{(i)} \in [0,x] \right) \right\}
\end{equation}
$X_{(0)}^{(i)}$ represents the order statistics for the particles. 
}
\end{itemize}
From these distributional estimates for density, distribution and quantile function, one can then obtain a particle based weighted dirac measure can estimate the risk measures for any $\alpha \in (0,1)$ by:
\begin{itemize}
\item[]{\textbf{Value-at-Risk (VaR): } directly obtained using the estimated quantile function!}
\item[]{\textbf{Expected Shortfall (ES): } estimated via
\begin{equation*}
\widehat{\mathrm{ES}_{Z}}(\alpha) \approx \frac{1}{N}\sum_{i=1}^N X^{(i)}_{0:n^{(i)}} W\left(X^{(i)}_{0:n^{(i)}}\right) \mathbb{I}\left[X^{(i)}_{0:n^{(i)}} \geq F^{\leftarrow}\left(1-\frac{1-\alpha}{\mathbb{E}[N]}\right) \right],
\end{equation*}
}
\item[]{\textbf{Spectral Risk (SRM): } the SRM for a weight function $\phi:[0,1] \mapsto \mathbb{R}$ is given by 
\begin{equation*}
\widehat{\mathrm{SRM}_{Z}}(\phi) = \frac{1}{N}\sum_{i=1}^N X^{(i)}_{(0):n^{(i)}} \phi\left(p_i\right)\triangle_{p_i} 
\end{equation*}
with $p_i = \sum_{j = 1:i} W\left(X^{(i)}_{(0):n^{(i)}}\right)$.
}
\end{itemize}
For additional discussions and detailed examples and applications of this numerical approach to risk estimation can be found in \cite{peters2007simulation} and \cite{shevchenko2011modeling}, and in financial asset pricing in \cite{doucet2010solving}. To conclude this illustrative section, we extend the previous example based on the popular Poisson-Log Normal LDA risk process model to include the path-space particle solution discussed above.

\begin{example}[Poisson-Log Normal LDA Model (continued)]{Consider the Poisson-Log Normal compound process detailed in Example 1.
We demonstrate results for standard Monte Carlo approach and compare results to the path-space particle solution discussed and detailed in Algorithm 2. The Monte Carlo solution generated N=50mil samples (hence effectively exact as Monte Carlo error is insignificant) and a grid based solution was adopted for the particle solution with N=50k per grid point giving a total of NT=500k, with a grid width = 1. The results for the estimation of the quantiles (rounded to integer) which includes the Basel II/III regulatory standard for Economic Capital and Regulatory Capital reporting are provided in the following table for two sets of parameter settings of $\lambda = 2$, $\mu = 2$ and $\sigma = 0.5$ and $\lambda = 2$, $\mu = 2$ and $\sigma = 1$. Two sets of parameters were considered, one based on the observed poor finite quantile performance of the Single Loss Approximate and the other in the setting in which the SLA is known to perform more accurately. The Monte Carlo errors for the Standard Monte Carlo approach are insignificant due to the massive sample size, hence we treat this as the exact solution upon which we compare the particle solution and the single loss approximation. The particle solution presents a 95\% confidence interval and the single loss approximation simply reports the asymptotic approximation no explicit error can be calculated for the point estimated quantile (as discussed above).
\begin{table*}[htbp]
	\centering
		\begin{tabular}{|c|c|c|c|c|c|c|} \hline
		\textbf{Quantile Level} & \multicolumn{2}{|c|}{\textbf{Standard Monte Carlo}} & \multicolumn{2}{|c|}{\textbf{Particle Solution (Alg. 2)}} & \multicolumn{2}{|c|}{\textbf{Single Loss Approximation}} \\ \hline \hline
        & $\sigma = 0.5$ & $\sigma = 1$ & $\sigma = 0.5$ & $\sigma = 1$ & $\sigma = 0.5$ & $\sigma = 1$\\ \hline		
50\% 		& 14 & 16 & 15 [14,16] & 16 [13,17] & 10 & 14\\
80\% 		& 27 & 39 & 25 [26,28] & 41 [39,43] & 14 & 26\\
90\% 		& 35 & 57 & 33 [31,35] & 55 [52,59] & 16 & 38\\
95\% 		& 42 & 77 & 40 [38,43] & 74 [70,79] & 19 & 52\\
99\% 		& 57 & 129 & 55 [54,56] & 123 [119,127] & 26 & 97\\
99.5\% 	& 77 & 234 & 73 [68,79] & 227 [218,240] & 38 & 198\\
99.95\% & 83 & 276 & 79 [73,91] & 270 [261,282] & 42 & 240\\ \hline		
		\end{tabular}
	\caption{Standard Monte Carlo Solution (exact) versus Particle Solution and First Order Single Loss Approximations.}
	\label{tab:MCvsPart}
\end{table*}
}
\end{example}
The results that we present in Table I are obtained on a linearly spaced grid of width 1. However, this can be changed to either include a non-linear spacing, placing more points around the mode and less points in the tails or as we detailed straightout evaluation on an interval, avoiding the discretization of the grid. For the sake of comparison between the standard Monte Carlo and the importance sampling estimates, we histogram the standard Monte Carlo procedure samples using unit length bins. We can see two things from Table I, firstly as expected the particle based solution performs accurately under any parameter settings for a modest computational budget. When compared to the Single Loss Approximation, we see that there is two clear advantages in having a complementary particle solution, since we obtain measures of uncertainty in the quantile point estimates, trivially. Secondly, we demonstrate that the Single Loss Approximations may not be as accurate as required for even these simple models at quantiles that may be of interest to assessment and are required for reporting of capital figures under financial regulation standards.

\bibliographystyle{ieeetr}
\bibliography{MCQMC.paper.references}
  
\end{document}